\documentclass[aps, pra, preprint, amsmath, amssymb, showpacs]{revtex4-2}
\usepackage{graphicx}
\usepackage{hyperref}

\begin{document}

\title{Polynomial propagators for classical molecular dynamics}
\author{Ivan Kondov}
\email{ivan.kondov@kit.edu}
\affiliation{Scientific Computing Center, Karlsruhe Institute of Technology, Hermann-von-Helmholtz-Platz 1,
76344 Eggenstein-Leopoldshafen, Germany}
\date{\today}

\begin{abstract}
Classical molecular dynamics simulation is performed mostly using the established velocity Verlet integrator or other symplectic propagation schemes. In this work, an alternative formulation of numerical propagators for classical molecular dynamics is introduced based on an expansion of the time evolution operator in series of Chebyshev and Newton polynomials. The suggested propagators have, in principle, arbitrary order of accuracy which can be controlled by the choice of expansion order after that the series is truncated. However, the expansion converges only after a minimum number of terms is included in the expansion and this number increases linearly with the time step size.
Measurements of the energy drift demonstrate the acceptable long-time stability of the polynomial propagators. It is shown that a system of interacting Lennard-Jones particles is tractable by the proposed technique and that the scaling with the expansion order is only polynomial while the scaling with the number of particles is the same as with the conventional velocity Verlet. The proposed method is, in principle, extendable for further interaction force fields and for integration with a thermostat, and can be parallelized to speed up the computation of every time step.
\end{abstract}


\maketitle

\newpage

\section{Introduction}
\label{sec:intro}

Molecular dynamics has become one of the most employed methods to simulate the classical motion of particles in complex models on the nanometer scale, such as biological systems and materials. Today's supercomputers allow the simulation of very large systems containing a hundred million particles \cite{chan2014ss:2503, mei2011sz:1}. While current computational methods treat the spacial scaling of molecular dynamics very efficiently, the computational scaling of the numerical integration in time is limited \cite{leim1996rs:161,hoch1998l:421,izag1999rs:9853}.

Commonly used schemes, such as the classical symplectic integrators, are commonly based on Trotter expansion of the time evolution operator~\cite{tuck1992bm:1990}. High-order symplectic integrators have been developed based on Baker-Campbell-Hausdorff (BCH) expansions~\cite{yosh1990:262, cham2000m:425}. Symplectic integrators have been reviewed extensively in the literature, e.g. in~Refs.~\citenum{mcla1992a:541, gray1994ns:4062}, and have the practical property of being energy conserving over a large number of integration steps. Among the various symplectic methods of different order, very common is the velocity Verlet algorithm \cite{verl1967:98} which is also simple to implement in computational codes.

Various methods have been developed to increase the time step and accelerate time integration in long-time simulations without affecting integration stability \cite{leim1996rs:161,hoch1998l:421,izag1999rs:9853}. This is particularly necessary if the time span, that the simulation is supposed to cover, is several orders of magnitudes longer that the time step. 
In the SHAKE \cite{ryck1977cb:327} and RATTLE \cite{ande1983:24} algorithms, selected atoms involved in fast quasi-periodic motion are constrained which allows choosing larger time steps. Using constraints with velocity-explicit algorithms has been source of numerical stability issues \cite{lipp2007bd:046101}. Other approaches still consider separation of time scales but do not impose constraints, e.g. the RESPA method \cite{tuck1992bm:1990}, or exploit time-scale difference in mixed quantum-classical dynamics by treating the quantum evolution separately with the Lanczos' method~\cite{hoch1999l:620}. A prerequisite to use these methods is that the time scales of the problem at hand must be separable.

The \emph{parallel-in-time} methods \cite{gand2015}, in particular the \texttt{parareal} algorithm \cite{baff2002bm:057701, audouze2009, lego2013ls:a1951, byla2013ww:074114, dai2013ll:717, gand2014h:2}, split the time integration grid into sections that are treated in parallel which speeds up the simulation with the number of time sections. This approach works only if at least the initial conditions at the beginning of each time section or a rough approximation of particles' trajectories in each time section are available. For example, the  \texttt{parareal} algorithm bases on two solvers with different accuracy: first the whole trajectory is sequentially computed with a less accurate integrator and then split into sections that are processed in parallel with an accurate integrator (corrector). Furthermore, accelerated integration methods for molecular dynamics have been developed and employed in biomolecular simulations, for example by applying semi-analytical integration \cite{mich2015d:336} and the mollified impulse method combined with co-rotational filters \cite{fath2017hs:180}.

The spectral deferred correction (SDC) methods \cite{dutt2000gr:241, spec2015re:843} are a class of methods to compute the higher-order collocation solution to an ordinary differential equations by successively applying correction sweeps using a low-order method, for example the velocity Verlet or even the first-order Euler method. The method has been applied to integrate the equations of motion for charged particles in electromagnetic fields using the velocity-Verlet-like Boris integrator as low-order method~\cite{wink2015sr:456}.  

In this paper, an alternative numerical scheme is proposed that may accelerate the simulation by allowing significantly larger time steps while maintaining stability and accuracy. The increased number of computations within every time step can be parallelized to accelerate the algorithm. If necessary, the trajectory can be generated at intermediate times within a time step with relatively small additional computational effort scaling only linearly with the number of degrees of freedom and with the number of approximating polynomial terms. 

The proposed scheme is based on expansion of the classical propagator in series of polynomials and outlined in Section~\ref{sec:classical-expansions}. This technique has been successfully applied to quantum dynamics \cite{tale1984k:3967, lefo1991bc:59, ashk1995kr:10005, nett1996hs:47, guo1999c:6626, chen1999g:19, xie2000cg:5263, schr2000kk:77, kond2001ks:1497, lubi2002:459, huis1998pk:29, huis1999pk:5538, berm92kt:1283} and yet not to classical dynamics. A direct application of the quantum dynamical polynomial scheme to classical dynamics is not possible because of the different representation and different spectral properties of the Liouville operator in classical systems, e.g. the numerical expression evaluations and an adequate choice of the spectral width. A detailed analysis of the procedure for a quantum mechanical system, including stability criterion and error estimates can be found in Ref.~\citenum{lubi2008:book}. While there is still need for further theoretical investigations of this specific mathematical problem for a classical system, here rather a practical/experimental numerical study is provided.

The suggested polynomial propagators are not symplectic, however, the error in energy conservation, i.e. the energy drift, can be fully controlled by the number of polynomial terms included in the expansion which will be shown in the numerical experiments in Section~\ref{subsec:1p}. 
In addition, the influence of the spectral width parameter and the long-time behavior are investigated in detail in Section~\ref{subsec:1p}.

The current study is only a proof of principle limited to toy model systems with a small number of particles without thermostat. The system Hamiltonian of the studied model systems is separable, i.e. $H(\mathbf{p}, \mathbf{q}) = T(\mathbf{p})+V(\mathbf{q})$, and quadratic in the momenta $\mathbf{p}$ that are limitations usually applying to the typical systems studies by molecular dynamics simulations in Cartesian coordinates. In addition, the potential $V(\mathbf{q})$ considered here includes only pair interactions that is not due to principal limitations of the scheme but for the sake of focusing on the time integration aspect. In Section~\ref{subsec:mp} it will be shown that the polynomial propagators have the same scaling with number of particles as the velocity Verlet integrator for any number of particles and for any expansion order. In the limit of many particles the scaling with the expansion order is polynomial.

\section{Expansions of the classical propagator}
\label{sec:classical-expansions}

The time evolution operator, which is also known as the classical propagator~\cite{tuck1992bm:1990}, is defined as $e^{i\mathcal{L}t}$, where $t$ is the time and $\mathcal{L}$ is the Liouville operator that for a many-particle system with Hamiltonian $H$ is defined as
\begin{equation}
i\mathcal{L} z
=\{z, H\}
= \sum\limits_{j=1}^M
\left(
\frac{\partial H}{\partial p_j}\frac{\partial z}{\partial q_j}
-\frac{\partial H}{\partial q_j}\frac{\partial z}{\partial p_j}
\right)
\label{eq:poisson}
\end{equation}
where the summation is over all degrees of freedom $M$. The variable $z$ can be any function in phase space, e.g.~the probability density $f(q_1, \dots q_M, p_1, \dots, p_M)$ or a phase space variable, such as any position $q_j$ or momentum $p_j$. The time evolution of $z$ is given by
\begin{equation}
z(t) = e^{i\mathcal{L} t} z(0)
\label{eq:time-evolution}
\end{equation}
which is the formal solution of the equation of motion
\begin{equation}
\frac{\partial z}{\partial t} = i\mathcal{L} z.
\label{eq:eom}
\end{equation}
When $z \in \{q_1, \dots q_M, p_1, \dots, p_M\}$ then Eq.~(\ref{eq:eom}) is a system of coupled ordinary differential equations. For a system with conservation of energy, $\mathcal{L}$ is a Hermitian operator in the space of square integrable functions of the positions $q_j$ and the momenta $p_j$ which implies that the eigenvalues of $\mathcal{L}$ are real.

\subsection{Chebyshev expansion}
\label{subsec:chebyshev}

The Chebyshev expansion of the classical propagator is closely related to the problem of computing the matrix exponential which has been solved and the numerical implementations have been shown to work efficiently and accurately \cite{shar2018s:613, moor2011:537, auck2010bh:359, berg2000v:27, tale1989:25}. The methods of Chebyshev and of Lanczos are known to provide efficient polynomial approximations of the matrix exponential (see e.g.~Ref.~\citenum{lubi2008:book} for an introduction). The approach used here is based on an expansion of $e^{Ab}$ into series of Chebyshev polynomials $\Psi_n(A)$:

\begin{equation}
e^{Ab} = \sum_{n=0}^\infty (2-\delta_{0n}) J_n(b) \Psi_n(A).
\label{eq:chebyshev-scalar}
\end{equation}
The expansion converges if $||A|| \le 1$, i.e. all eigenvalues of $A$ are bounded between $-1$ and $1$. In the case of dynamics the expansion (\ref{eq:chebyshev-scalar}) is applied to the propagator in Eq.~(\ref{eq:time-evolution}). The matrix $A$ has to be replaced with the operator $i\mathcal{L}$ and the scalar parameter $b$ has to be replaced with the time $t$. In practice, the exact infinite series is truncated after some order $n=N$.

Chebyshev expansion of quantum dynamics propagators has been shown to yield very stable and accurate numerical schemes \cite{tale1984k:3967, lefo1991bc:59, ashk1995kr:10005, nett1996hs:47, guo1999c:6626, chen1999g:19, xie2000cg:5263, schr2000kk:77, kond2001ks:1497, lubi2002:459}.  In the case of quantum dynamics, this approach relies on the matrix representation of the Liouville operator and has two major benefits: i)~The Chebyshev polynomials in the expansion are evaluated using plain matrix--matrix and matrix--vector multiplications that have straightforward numerical implementation based on the Clenshaw algorithm (see e.g.~Ref.~\citenum{lubi2008:book}). In contrast, the expansion of the classical propagator requires sophisticated symbolical evaluations followed by numerical evaluations which, in general, scale exponentially with the order of the truncated expansion~$N$. ii)~The spectral bounds of a Hermitian matrix, necessary for the Chebyshev expansion, can be computed straightforwardly using standard linear algebra numerical methods, e.g.~the power iteration. In contrast, calculating the spectrum of the classical Liouville operator is a non-trivial problem that has been addressed in the literature \cite{wilk1997b:27, carv1996a:3597, kumi1984:821, zwan1966:170}.

The Chebyshev expansion scheme is stable when the spectrum of the Liouville operator is bounded~\cite{lubi2008:book}, and is either pure imaginary or pure real, or complex with small real components, or complex with small imaginary components \cite{ashk1995kr:10005, guo1999c:6626}. The Chebyshev scheme becomes unstable when the spectrum of the Liouville operator is extended along both the real and the imaginary axes of the complex plane~\cite{midg2000w:920}. Because the classical Liouville operator $\mathcal{L}$ is Hermitian, its spectrum is real (thus the eigenvalues of $i\mathcal{L}$ are imaginary) which makes an expansion in Chebyshev polynomials especially suitable.

Another stability condition for the Chebyshev expansion is that all eigenvalues are either in the range $[-1, 1]$ or in $[-i, i]$. For this reason the Liouville operator has to be scaled and shifted as $\tilde{\mathcal{L}} = 2(\mathcal{L} - L_0)/\Delta L$ in order to restrict its eigenvalues in one of these ranges. Therefore, the lower and the upper bounds $L_{\mathrm{min}}$ and $L_{\mathrm{max}}$ of the $\mathcal{L}$ spectrum have to be known. Generally it is not possible to find the largest and the lowest eigenvalues of $\mathcal{L}$. Then the half span $\Delta L/2 = (L_{\mathrm{max}} - L_{\mathrm{min}})/2$ and the middle point $L_0 = L_{\mathrm{min}} + \Delta L/2$ of the spectrum can be regarded as adjustable parameters. Previously it has been shown that the spectrum of the Liouville operator is always symmetric with respect to zero \cite{spoh1975:323}. For this reason, it is assumed that $L_0=0$ in the following without loss of generality.
The Chebyshev expansion for the thus scaled Liouville operator, truncated after order $N$, will read
\begin{equation}
z(t) \approx \sum_{n=0}^N a_n(\alpha) \tilde{\Phi}_n(i\tilde{\mathcal{L}} z(0)) 
\label{eq:chebyshev}
\end{equation}
where $\alpha=t\Delta L/2$ and $\tilde{\Phi}_n(ix) = i^n \tilde{\Psi}_n(x)$ are the scaled modified Chebyshev polynomials~\cite{tale1986:11} (see Appendix~\ref{appendix-chebyshev}). The expansion coefficients are $a_n(\alpha) = (2-\delta_{n0})J_n(\alpha)$ with $J_n(\alpha)$ the Bessel functions of the first kind. A mathematically equivalent expansion of the propagator $e^{i H t}$, i.e. with the Hamiltonian instead of the Liouville operator, has been performed in the context of quantum dynamics \cite{tale1984k:3967, kosl1986t:223, lefo1991bc:59}. The scaled modified Chebyshev polynomials $\tilde{\Phi}_n(i\tilde{\mathcal{L}} z)$ are computed using a recurrence relation \cite{tale1984k:3967, kosl1986t:223, lefo1991bc:59}:
\begin{equation}
\tilde{\Phi}_{n+1}(i\tilde{\mathcal{L}} z) = 
2i\tilde{\mathcal{L}}\tilde{\Phi}_n(i\tilde{\mathcal{L}} z) + \tilde{\Phi}_{n-1} (i\tilde{\mathcal{L}} z).
\label{eq:chebyshev-rec}
\end{equation}
The recursion is started with $\tilde{\Phi}_0(i\tilde{\mathcal{L}} z) = z$ and $\tilde{\Phi}_1(i\tilde{\mathcal{L}} z) = i\tilde{\mathcal{L}} z$. Before starting the time propagation, the $\tilde{\Phi}_n(i\tilde{\mathcal{L}} z)$ are evaluated symbolically in advance using the recursion formula (\ref{eq:chebyshev-rec}) and stored in memory as expressions.

The time integration is carried out on a discrete time grid $\{t_k\}$ with $k=0, 1, \dots$ equally spaced by a time step with size $\Delta t = t_{k+1} - t_k$. To compute $z(t_{k+1})$ the vector $\tilde{\Phi}_n(i\tilde{\mathcal{L}} z)$ is numerically evaluated with the current $z(t_k)$. Because the vector $\tilde{\Phi}_n(i\tilde{\mathcal{L}} z)$ does not depend explicitly on time, it is possible to compute $z(t)$ at any time between $t_k$ and $t_{k+1}$ without recomputing the vector. Only the expansion coefficients $a_n(\alpha)$ with $\alpha = \Delta t \Delta L / 2$ have to be computed and stored in advance for different time step sizes. Such additional calculations consist of a single inner product with computational cost scaling only linearly with the vector length of $\tilde{\Phi}_n(i\tilde{\mathcal{L}} z)$. It is known that Chebyshev expansion has uniform accuracy in the range $[-1, 1]$ and thus the accuracy of the Chebyshev polynomial propagator will be uniform over the whole spectrum of the Liouville operator provided that $\mathcal{L}$ is properly scaled \cite{lefo1991bc:59, guo1999c:6626}. Therefore this propagation method is especially suitable for problems without separation of scales and poses no principal limitations on the choice of the time step size. In particular, the expansion (\ref{eq:chebyshev}) converges exponentially for $N>\alpha$ \cite{tale1984k:3967, lefo1991bc:59, ashk1995kr:10005} due to the properties of the Bessel functions $J_n(\alpha)$. The only price to pay is the proportional increase of the minimum $N$ necessary for convergence with increasing the time step.

\subsection{Newton polynomial interpolation}
\label{subsec:newton}

By analogy with the quantum mechanical propagator \cite{huis1999pk:5538, huis1998pk:29, ashk1995kr:10005, berm92kt:1283}, the classical propagator in Eq.~(\ref{eq:time-evolution}) can be approximated by interpolation with a Newton polynomial
\begin{equation}
e^{i\mathcal{L}t} z \approx \sum_{n=0}^N D_n(t) P_n(i\mathcal{L} z)  = \sum_{n=0}^N D_n(t) \prod_{k=0}^{n-1} (i\mathcal{L}-\lambda_k) z
\label{eq:newton}
\end{equation}
where $P_n(i\mathcal{L})$ are the Newton basis polynomials. The coefficients $D_n(t)$ are the $n$th divided differences of the function $f(\lambda)=e^{\lambda t}$ with argument $\lambda$ and parameter $t$ at the interpolation points $\{\lambda_0, \dots, \lambda_{N-1}\}$: $D_0 =  f[\lambda_0]$, $D_1 = f[\lambda_0, \lambda_1]$, ..., $D_k = f[\lambda_0, \dots, \lambda_k]$. The divided differences are calculated recursively based on their definition:
\begin{subequations}
\begin{eqnarray}
f[\lambda_k] & = & f(\lambda_k) \\
f[\lambda_{k-1}, \lambda_k] & = &\frac{f[\lambda_k]-f[\lambda_{k-1}]}{\lambda_k-\lambda_{k-1}} \\ \nonumber
& \dots & \\
f[\lambda_l, \dots, \lambda_k] & = & \frac{f[\lambda_{l+1}, \dots, \lambda_k]-f[\lambda_l, \dots, \lambda_{k-1}]}{\lambda_k-\lambda_l}~.
\label{eq:dds}
\end{eqnarray}
\end{subequations}
The interpolation points, also called Leja points \cite{breu2018kv:279, reic1990:332, berm92kt:1283, ashk1995kr:10005, bagl1998cr:124, huis1998pk:29, huis1999pk:5538}, are here determined using the algorithm by Ashkenazi et al.~\cite{ashk1995kr:10005} described briefly in the following. An initial domain is chosen so that its capacity is close to unity to avoid overflows. The interpolation points $\{\lambda_k\}$ are chosen from a set of trial points uniformly distributed in the initial domain in such a way to maximize the denominators of the divided differences. The chosen interpolation points are scaled by $\rho =  \sqrt[N]{\prod_{j=0}^{N-1}|\lambda_c - \lambda_j|}$, i.e. $\tilde{\lambda}_k = \lambda_k/\rho$. This adjustment is necessary to avoid overflow or underflow in the numerical evaluation of the product on the right hand side of Eq.~(\ref{eq:newton}). Additionally, the Liouville operator is scaled and shifted, so that all its eigenvalues are inside the scaled interpolation domain, by making the following substitutions in Eq.~(\ref{eq:newton}):
$\{\lambda_k\} \rightarrow \{\tilde{\lambda}_k\}$, $\mathcal{L} \rightarrow \tilde{\mathcal{L}} = (\mathcal{L}-L_0)/\sigma$, $D_n(\{\tilde{\lambda}_k\},\ t) \rightarrow \tilde{D}_n(\{\sigma\tilde{\lambda}_k+L_0\},\ t)$, $\sigma = \Delta L\rho$, where $L_0$ and $\Delta L$ are the center and the width of the $\mathcal{L}$ spectrum, respectively. Because the spectrum of the Liouville operator is symmetric with respect to zero \cite{spoh1975:323} it is assumed that $L_0=0$ and no shift is performed in the following.

Before the propagation in time begins, the scaled basis polynomials $\tilde{P}_n(i\tilde{\mathcal{L}} z)$ are evaluated by carrying out the product on the right hand side of Eq.~(\ref{eq:newton}) recursively by the formula
\begin{equation}
\tilde{P}_{n}(i \tilde{\mathcal{L}} z) = 
(i \tilde{\mathcal{L}} - \tilde{\lambda}_{n-1}) \tilde{P}_{n-1} (i \tilde{\mathcal{L}} z)
\label{eq:newton-rec}
\end{equation}
where the recursion starts with $\tilde{P}_0(i \tilde{\mathcal{L}} z) = z$ and $i \tilde{\mathcal{L}} \tilde{P}_n$ is evaluated according to Eq.~(\ref{eq:poisson}) as it will be outlined in Section~\ref{subsec:liouville-powers}.

The arithmetic operations in the scaled version of Eq.~(\ref{eq:newton}) are performed numerically on a grid of equally spaced time points $\{t_k\}$ to get the numerical solution $z(t_k)$. Because the vector $\tilde{P}_n(i\tilde{\mathcal{L}} z)$ is not explicitly time-dependent, it is possible to compute $z(t)$ at any time between two arbitrary time grid points $t_k$  and $t_{k+1}$ without recomputing the vector. Only the divided differences coefficients $\tilde{D}_n(\Delta t)$ have to be computed and stored in advance for all needed time step sizes. The necessary additional calculations are the multiplications and the summation in Eq.~(\ref{eq:newton}) and therefore scale only linearly with the length of the vector.

Similarly to the Chebyshev propagator, the Newton polynomial propagator has uniform accuracy over the whole spectral range of $\mathcal{L}$ \cite{kosl1994:145} provided that $\mathcal{L}$ is properly scaled as described above.

\subsection{Expansion in Krylov subspace}
\label{subsec:krylov}

In the following, an alternative strategy for expansion of the propagator $e^{i\mathcal{L}t}$ is presented which is based on the Krylov subspace. In the following, the scaled versions of all quantities, defined in Sections~\ref{subsec:chebyshev} and \ref{subsec:newton}, will be used but the tilde (\textasciitilde) will be omitted for the sake of readability. The elements of the Krylov subspace are generated by successive application of the Liouville operator to a dynamic variable $z: \{z, i\mathcal{L} z, (i\mathcal{L})^2 z, \dots, (i\mathcal{L})^N z\}$, where $N$ is the dimension of the Krylov subspace. Then the time evolution of $z$ can be approximated as
\begin{equation}
z(t) = e^{i\mathcal{L}t} z(0) \approx \sum_{n=0}^N K_n(t) (i\mathcal{L})^n z(0).
\label{eq:krylov}
\end{equation}
Depending on the choice of the expansion coefficients $K_n(t)$ various propagation methods can be derived. For example, the choice $K_n(t)=t^n/n!$, corresponding to a Taylor expansion of $e^{i\mathcal{L} t}$ in time around $t=0$, yields the commonly used Euler method (for $N=1$) and the Runge-Kutta method (e.g. for $N=4$). The Taylor expansion is non-uniform, i.e. the error becomes large for larger eigenvalues of $\mathcal{L}$ unless the time step is adaptively reduced. This is the reason why the Runge-Kutta method is generally limited to small time steps determined by the largest eigenvalue of $\mathcal{L}$. 

The expansion in Krylov subspace suggests an alternative strategy to symbolically evaluate the expansion terms in Eqs.~(\ref{eq:chebyshev}) and (\ref{eq:newton}) without having to use the scheme-specific recurrence relations Eqs.~(\ref{eq:chebyshev-rec}) and (\ref{eq:newton-rec}). Particularly, it can be shown that Chebyshev and Newton polynomial propagators can also be represented as Eq.~(\ref{eq:krylov}) whereby $K_n(t)$ become scheme-specific and the problem is reduced to a recursive evaluation of the monomials $(i\mathcal{L})^n z$. This is achieved when  the scaled Chebyshev polynomials $\Phi_n(i \mathcal{L}) z$  in Eq.~(\ref{eq:chebyshev-rec}) are rewritten into their explicit form
\begin{equation}
\Phi_n(i \mathcal{L}) z = \sum_{k=0}^n C_k^{(n)}(i \mathcal{L})^k z
\label{eq:chebyshev-nonrec1}
\end{equation}
where the coefficients $C_k^{(n)} $ form a triangle with the formula \cite{oeis:A081265}
\begin{equation}
C_k^{(n)} =
\begin{cases}
0 & \text{if}\ n+k\ \text{is odd} \\
1 & \text{if}\ k = 0\ \text{and}\ n\ \text{is even} \\
2^{k-1} \frac{n}{k}  {\frac{n+k}{2}-1 \choose k-1} & \text{otherwise}.
\end{cases}
\label{eq:chebyshev-nonrec2}
\end{equation}
Analogously, the Newton basis polynomials $P_n(i\mathcal{L}z)$ can be represented as
\begin{equation}
P_n(i\mathcal{L}z)
= \prod_{k=0}^{n-1} (i\mathcal{L}-\lambda_k) z
= \sum_{k=0}^n C^{(n)}_k (i\mathcal{L})^{n-k} z~.
\label{eq:newton-nonrec1}
\end{equation}
The expansion coefficients can be derived by induction:
\begin{equation}
C^{(n)}_k = (-1)^k \sum_{c \in S} \prod_{x \in c} x = (-1)^k \sum_{l=1}^{n \choose k} \left[\lambda_0 \lambda_1 \dots \lambda_{n-1}\right]_k
\label{eq:newton-nonrec2}
\end{equation}
where $S$ is the set of all $k$-combinations of the set of the first $n$ interpolation points $\{\lambda_0, \dots, \lambda_{n-1}\}$. An alternative way to compute $C^{(n)}_k$ is to expand the polynomial $\prod_{j=0}^{n-1} (1-\lambda_j x)=\sum_{k=0}^n C^{(n)}_k x^k$.

\subsection{Evaluation of the basis vectors}
\label{subsec:liouville-powers}

The expansions in Eqs.~(\ref{eq:chebyshev-nonrec1}) and (\ref{eq:newton-nonrec1}) are equivalent to Eqs.~(\ref{eq:chebyshev-rec}) and (\ref{eq:newton-rec}), respectively. Eqs.~(\ref{eq:chebyshev-rec}) and (\ref{eq:newton-rec}) require the evaluation of the Chebyshev and the Newton basis polynomials of $i\mathcal{L}$, respectively, whereas Eqs.~(\ref{eq:chebyshev-nonrec1}) and (\ref{eq:newton-nonrec1}) require evaluating the Krylov subspace basis vectors, i.e. the monomials $(i\mathcal{L})^n$. Therefore, the two approaches yield expressions implying different number of symbolic evaluations and number of numeric operations during propagation. Depending on the specific interaction potentials, either the former or the latter expressions may have performance benefits. In the following, the evaluation of the monomials $(i\mathcal{L})^n$ will be outlined. As introduced above, it is assumed that the potential energy  $V$ depends only on the positions and the kinetic energy $T$ depends only on the momenta, i.e. the  Hamiltonian reads $H(q_1, \dots, q_M, p_1, \dots, p_M) = V(q_1, \dots, q_M) + T(p_1, \dots, p_M)$. The following compact notation is introduced:
\begin{equation}
a_j=\frac{\partial}{\partial q_j},\quad
b_j=\frac{\partial}{\partial p_j},\quad
A_j^{(1)} = \frac{\partial V}{\partial q_j},\quad
B_j^{(1)} = \frac{\partial T}{\partial p_j}
\end{equation}
where $q_j$ is the $j$th Cartesian position and $p_j$ is the pertinent momentum. Using the above notation, the Poisson bracket in Eq.~(\ref{eq:poisson}) can now be applied to a variable $z_l$ (e.g.~$q_l$ or $p_l$):
\begin{equation}
i\mathcal{L} z_l = \sum\limits_{\mu=1}^M \left(B_\mu^{(1)} a_\mu - A_\mu^{(1)} b_\mu\right) z_l~.
\label{eq:mpliouville}
\end{equation}
Using the definition in Eq.~(\ref{eq:mpliouville}) and that $T(p_1, \dots, p_M) = \sum_{j=1}^M p_j^2/(2 m_j)$ in Cartesian coordinates (with $m_j$ the mass pertinent to the $j$th degree of freedom), the operator $i\mathcal{L}$ is recursively applied symbolically as many times as necessary to compute $(i\mathcal{L})^n z_l$, i.e.
\begin{subequations}
\begin{eqnarray}
(i\mathcal{L})^n z_l & = & \prod\limits_{k=1}^n\sum\limits_{\nu_k=1}^M \left(B_{\nu_k}^{(1)} a_{\nu_k} - A_{\nu_k}^{(1)} b_{\nu_k}\right) z_l \\
& = & \mathcal{K}_l^{(n)}(B_1^{(1)}, \dots, B_M^{(1)}, B_1^{(2)}, \dots, B_M^{(2)}, \{A_{\nu_1}^{(1)}, A_{\nu_1, \nu_2}^{(2)}\dots, A_{\nu_1, \dots, \nu_n}^{(n)}\})
\label{eq:mpliouville-taylor}
\end{eqnarray}
\end{subequations}
where $\mathcal{K}_l^{(n)}$ are multivariate arithmetic polynomial expressions with the variables denoted in the brackets. For illustration, the explicit forms of these expressions for the one-particle case are derived in Appendix~\ref{appendix-one-dimension}. In the practical implementation used in the numerical experiments below the expressions are evaluated using a system for symbolic computation and then compiled in a just-in-time fashion. In Eq.~(\ref{eq:mpliouville-taylor}) the variables $B_j^{(1)} = p_j/m_j$, $B_j^{(2)} = 1/m_j$ for $j=1, \dots, M$ are the velocities and the inverted masses respectively, and $A^{(n)}_{\nu_1, \dots, \nu_n}$ are the partial derivatives of the potential energy with respect to the Cartesian positions
\begin{equation}
A^{(n)}_{\nu_1, \dots, \nu_n} = \frac{\partial^n V(q_1, \dots, q_M)}{\partial q_{\nu_1} \dots \partial q_{\nu_n}}
\label{eq:potential-energy-cartesian-derivs}
\end{equation}
with $\{\nu_1, \dots, \nu_n\}$ being $n$-combinations with repetition of $M$ degrees of freedom. In general,  the quantities $A^{(n)}_{\nu_1, \dots, \nu_n}$ form $n$th-order tensors. In the special cases for $n=1$ and $n=2$,  the $A_{j}^{(1)}$ and $A_{j,l}^{(2)}$  are the elements of the gradient vector and of the Hessian matrix, respectively. 
Analogous expressions to Eq.~(\ref{eq:mpliouville-taylor}) can also be derived for the Chebyshev polynomials
\begin{subequations}
\begin{eqnarray}
\Phi_n(i\mathcal{L} z_l) & = & 2 \sum\limits_{j=1}^M \left(B_{j}^{(1)} a_{j} - A_{j}^{(1)} b_{j}\right)  \Phi_{n-1}(i\mathcal{L} z_l) + \Phi_{n-2}(i\mathcal{L} z_l) \\
& = & \mathcal{F}_l^{(n)}(B_1^{(1)}, \dots, B_M^{(1)}, B_1^{(2)}, \dots, B_M^{(2)}, \{A_{\nu_1}^{(1)}, A_{\nu_1, \nu_2}^{(2)}\dots, A_{\nu_1, \dots, \nu_n}^{(n)}\})
\label{eq:mpliouville-chebyshev}
\end{eqnarray}
\end{subequations}
for $n=2, \dots, N$ and for the Newton basis polynomials
\begin{subequations}
\begin{eqnarray}
P_{n}(i\mathcal{L} z_l) & = & \sum\limits_{j=1}^M \left(B_{j}^{(1)} a_{j} - A_{j}^{(1)} b_{j}\right)P_{n-1} (i\mathcal{L} z_l) - \lambda_{n-1} P_{n-1} (i\mathcal{L} z_l) \\
& = &
\mathcal{P}_l^{(n)}(B_1^{(1)}, \dots, B_M^{(1)}, B_1^{(2)}, \dots, B_M^{(2)}, \{A_{\nu_1}^{(1)}, A_{\nu_1, \nu_2}^{(2)}\dots, A_{\nu_1, \dots, \nu_n}^{(n)}\})
\label{eq:mpliouville-newton}
\end{eqnarray}
\end{subequations}
for $n=1, \dots, N$. Note that the expressions (\ref{eq:mpliouville-taylor}), (\ref{eq:mpliouville-chebyshev}) and (\ref{eq:mpliouville-newton}) include derivatives of the kinetic energy only up to the second order and all higher derivatives are vanishing. 

The classical force fields used in molecular dynamics are commonly expressed in internal coordinates, such as Z-matrix coordinates. In such coordinates, every potential energy term depends on a single internal coordinate $Q_\alpha$:
\begin{equation}
V(Q_1, \dots, Q_{M_\mathrm{int}}) = \sum_{\alpha=1}^{M_\mathrm{int}}V_\alpha(Q_\alpha(q_{\mu_{\alpha, 1}}, \dots, q_{\mu_{\alpha, M_\alpha}})).
\label{eq:potential-energy-internal}
\end{equation}
Every $Q_\alpha$ depends on only a small subset of Cartesian coordinates $\{q_{\mu_{\alpha, 1}}, \dots, q_{\mu_{\alpha, M_\alpha}}\} \subseteq \{q_1, \dots, q_M\}$ whereby  $M_\alpha \ll M$ in large many-particle systems.
Differentiating Eq.~(\ref{eq:potential-energy-internal}) with respect to the Cartesian coordinates up to the $n$th order 
\begin{equation}
A^{(n)}_{\nu_1, \dots, \nu_n} = \sum_{\alpha=1}^{M_\mathrm{int}}
\frac{\partial^n V_\alpha(Q_\alpha(q_{\mu_{\alpha, 1}}, \dots, q_{\mu_{\alpha, M_\alpha}}))} {\partial q_{\nu_1} \dots \partial q_{\nu_n}}
\label{eq:potential-energy-internal-derivs}
\end{equation}
yields, compared to Eq.~(\ref{eq:potential-energy-cartesian-derivs}), a strongly reduced number of Cartesian derivatives
\begin{equation}
\frac{\partial^n Q_\alpha(q_{\mu_{\alpha, 1}}, \dots, q_{\mu_{\alpha, M_\alpha}})}{\partial q_{\nu_1}\dots \partial q_{\nu_n}}
\label{eq:int-cart-derivs}
\end{equation}
where $\{\nu_1, \dots, \nu_n\}$ are $n$-combinations with repetition from the subset $\{q_{\mu_{\alpha, 1}}, \dots, q_{\mu_{\alpha, M_\alpha}}\}$.
The full derivatives on the right hand side of Eq.~(\ref{eq:potential-energy-internal-derivs}) can be evaluated using a generalized chain formula for a multivariate composite function~\cite{ma2009:n21, hard2006:r1}. In this work, an alternative approach is adopted using a system for symbolic computation outlined in~Section~\ref{subsec:workflow} below.

In the following, a computational complexity analysis will be provided. The total number of Cartesian derivatives (\ref{eq:int-cart-derivs}) up to order $N$ is given by
\begin{equation}
\sum\limits_{n=1}^{N}\sum\limits_{\alpha=1}^{M_\mathrm{int}}{M_\alpha+n-1\choose M_\alpha-1}.
\label{eq:number-of-derivatives}
\end{equation}
In practice, there are only a few internal coordinates with different symbolic expressions for which the derivatives have to be evaluated symbolically. For example, if the potential energy includes only pair interactions, then the derivatives for only one internal coordinate, i.e. the inter-particle distance, have to be  evaluated symbolically.

In the numerical experiments in Section~\ref{sec:measurements}, the method will be applied to model systems described by potentials for which correlations giving rise to mixed derivatives occur in pairs of particles. Such potentials can be the Lennard-Jones, Coulomb or gravitational potentials. For such potentials, Eq.~(\ref{eq:potential-energy-internal}) takes the form
\begin{equation}
V(q_{1x}, q_{1y}, \dots, q_{Nx}, q_{Ny}, q_{Nz}) = \sum\limits_{j<k} v(q_{jx}, q_{jy}, q_{kz}, q_{kx}, q_{ky}, q_{kz}) = \sum\limits_{j<k} v(r_{jk})
\label{eq:pair-potential-energy}
\end{equation}
where the internal coordinates $r_{jk}$ are the elements of the distance matrix
\begin{equation}
r_{jk} = | \mathbf{r}_j - \mathbf{r}_k | =  \sqrt{(q_{jx}-q_{kx})^2+(q_{jy}-q_{ky})^2+(q_{jz}-q_{kz})^2}~.
\label{eq:distance-matrix}
\end{equation}
Replacing $M_\alpha=6$ in Eq.~(\ref{eq:number-of-derivatives}) and omitting the sum due to identical derivatives expressions for different $\alpha$, the total number of Cartesian derivatives (\ref{eq:int-cart-derivs}) that have to be evaluated symbolically is reduced to 
\begin{equation}
\sum\limits_{n=1}^{N}{n+5\choose n} = 
\sum\limits_{n=1}^{N}{n+5\choose 5} \propto \mathcal{O}(N^6) .
\end{equation}
The number of derivatives can be further reduced by examining the potential energy terms for vanishing or identical derivatives. In the case of pair interaction potentials, the number of numerical evaluations of the expressions (\ref{eq:mpliouville-taylor}), (\ref{eq:mpliouville-chebyshev}) and (\ref{eq:mpliouville-newton}) scales only quadratically, i.e. $\mathcal{O}(M^2)$, with the number of particles. This scaling can be further reduced to $\mathcal{O}(M \log M)$ or $\mathcal{O}(M)$ using elaborated potential truncation schemes.

\subsection{Computational workflow}
\label{subsec:workflow}

\begin{figure}
\begin{centering}
\includegraphics[width=0.9\textwidth]{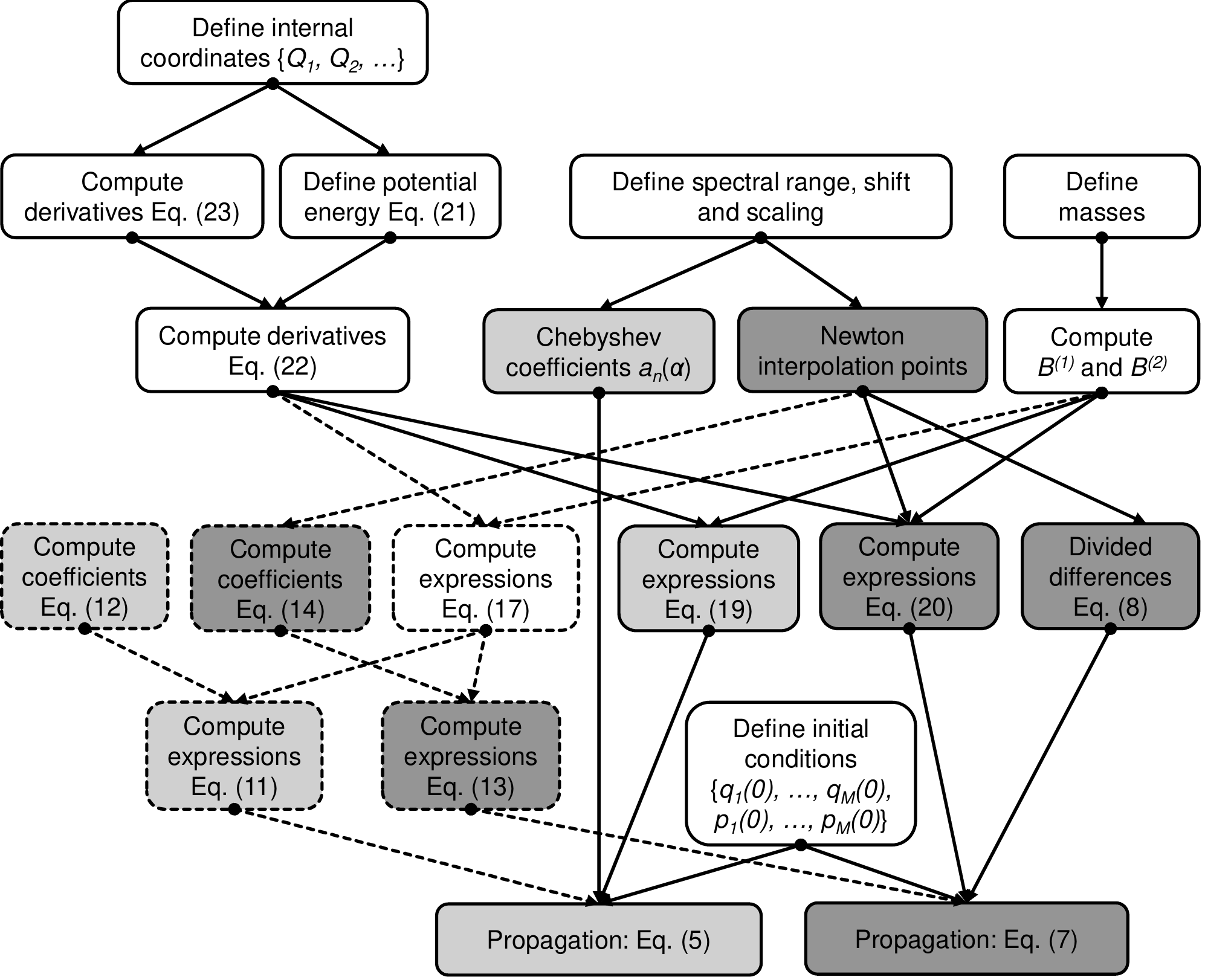}
\caption{Workflow for constructing and using the Chebyshev and Newton polynomial propagators drawn in light grey and dark grey, respectively. The white boxes contain steps that are used in both polynomial methods and dashed lines denote alternative routes based on the monomials $(i\mathcal{L})^n$ (cf.~Section~\ref{subsec:krylov}).}
\label{fig:workflow}
\end{centering}
\end{figure}

The overall computational workflow is graphed in Figure~\ref{fig:workflow}. To compute the expressions $(i\mathcal{L})^n z_l$, $\Phi_n(i\mathcal{L}z_l)$ and  $P_n(i\mathcal{L}z_l)$ in the first stage of the workflow, the Cartesian partial derivatives have been evaluated symbolically according to Eq.~(\ref{eq:potential-energy-internal-derivs})  and substituted in Eqs.~(\ref{eq:mpliouville-taylor}), (\ref{eq:mpliouville-chebyshev}) and (\ref{eq:mpliouville-newton}), respectively, using the system for symbolic computation \texttt{SymPy}~\cite{matt2017tr:e103}. 
In the second stage, before starting the propagation, the final expressions have been compiled into highly optimized numerical kernel functions, whereby the packages \texttt{NumPy}~\cite{numpy}, \texttt{Theano}~\cite{theano-full} and \texttt{TensorFlow}~\cite{abad2016bc:265} have been used as backends. The use of \texttt{Theano} or \texttt{TensorFlow} allows the execution on heterogeneous computing resources including GPUs.
In the third propagation stage, the inter-particle distance matrix (\ref{eq:distance-matrix}) has been computed using standard fast methods and used as input to evaluate the expressions from Eqs.~(\ref{eq:mpliouville-taylor}), (\ref{eq:mpliouville-chebyshev}) and (\ref{eq:mpliouville-newton}) numerically. The latter enter the propagator expansions Eqs.~(\ref{eq:chebyshev}) and (\ref{eq:newton}) that have been performed numerically as scalar products. 

The suggested computation strategy has some advantages compared to a traditional approach based on implementing explicit analytic expressions, e.g.~the forces of different force fields. The method can be employed with any force fields that have smooth high derivatives. As of now the method is realized in the important case of pair interaction force fields, such as Lennard-Jones and Coulomb interactions, because of the low implementation effort.  Furthermore, the numerical expressions evaluation in every propagation time step can be readily parallelized through splitting the expressions into chunks and distributing the chunks over the available computing elements. Finally, the computation of the chunks can be optimized and vectorized with the generation of the kernel functions.


\section{Numerical measurements}
\label{sec:measurements}

In the following, numerical experiments will be presented in the cases of one-particle and many-particle dynamics. The energy drift has been used as an accuracy measure for several reasons: To calculate the root mean square deviation (RMSD) of the positions or momenta, a very accurate reference solution of the equation of motion is necessary. The analytical solution can be used as reference but it is mostly not available, particularly for many-particle systems. Other error measures, e.g. the relative deviation from the reference solution, may not be finite at all times of integration for one-particle systems that are used in the following as test models. The energy drift has been regarded as criterion for long-term stability of time integration in molecular dynamics simulations \cite{cott2007t:507}. The energy drift $\Delta E_k$ can be computed without external reference using only the trajectories, i.e. the sets $\{q_j(t_k)\}$ and $\{p_j(t_k)\}$ for $j=1, \dots, M$,
\begin{equation}
\Delta E_k = \left|\frac{E_0 - E_k}{E_0}\right| = \left|\frac{E_0 - V(q_1(t_k), \dots, q_M(t_k))
-T(p_1(t_k), \dots, p_M(t_k))}{E_0}\right|
\label{eq:edrift}
\end{equation}
where $E_0$ is the total energy of the physical system, which is known and must be constant for the microcanonical ensemble considered in this paper, and $E_k$ is the energy at the discrete times $t_k$ at which the numerical integration has been performed. It is noted that the potential energy can always be shifted so that $E_0 \neq 0$. 

For comparison, a simple implementation of the velocity Verlet integrator, defined as
\begin{subequations}
\begin{eqnarray}
p_j(t_k+\Delta t/2) & := & p_j(t_k) - \frac{\Delta t}{2} A^{(1)}_j(q_1(t_k), \dots, q_M(t_k)) \\
q_j(t_{k+1}) & := & q_j(t_k) + \Delta t \frac{1}{m_j} p_j(t_k+\Delta t/2)  \\
p_j(t_{k+1}) & := & p_j(t_k+\Delta t/2) - \frac{\Delta t}{2} A^{(1)}_j(q_1(t_{k+1}), \dots, q_M(t_{k+1})) 
\label{eq:velocity-verlet}
\end{eqnarray}
\end{subequations}
has been used, where $-A^{(1)}_j(q_1, \dots, q_M)$ is the force (cf.~Eqs.~(\ref{eq:potential-energy-cartesian-derivs}) and (\ref{eq:potential-energy-internal-derivs})) and $\Delta t = t_{k+1}-t_k$ is the time step size.

The code has been implemented in Python 3.6 and made available as open source \footnote{The code is available under \url{https://gitlab.kit.edu/ivan.kondov/rotagaporp-c} under an open source license.}. Every simulation has been performed on one core of an Intel(R) Xeon(R) CPU E5-2670 processor at 2.60 GHz.

\subsection{One-particle system: classical Morse oscillator}
\label{subsec:1p}

For a general validation the implemented polynomial propagators have been first tested with the classical harmonic potential, constant-force potential (ballistic motion) and Morse potentials comparing the position $q$ and momentum $p$ of the particle to the analytic solutions of the equations of motion. In addition, the numerical solutions by the polynomial propagators for anharmonic double-well and Lennard-Jones potentials were validated against the solutions obtained with the velocity Verlet integrator. 

In the following, only measurement results for a particle in one spacial dimension in Morse potential will be discussed. The Morse potential
\begin{equation}
V(q) = D \left[e^{-2 \kappa (q-q_0)} - 2 e^{-\kappa (q-q_0)}\right]
\label{eq:morse}
\end{equation}
is an effective potential used to describe the dynamics of a covalent chemical bond in a diatomic molecule. $D$ is the dissociation energy, $q_0$ is the equilibrium bond length and $\kappa$ describes the spatial extent (the ``softness'') of the bond. The high-order derivatives of the Morse potential have closed analytic forms.
The parameters used in the simulations are summarized in Table~\ref{tab:morse-params}.

\begin{table}[b]
\caption{
  Morse potential and integrator parameters that were constant throughout this study if not specified otherwise.
  \label{tab:morse-params}
}
\begin{center}
\begin{tabular}{rrrrrrrr}
\hline\hline
$D$ & $\kappa$ & $q_0$  & $t_0$ & $t_\mathrm{end}$ & $q(t_0)$ & $p(t_0)$ & $\Delta L$\\ \hline
 1 & 1 & 1 &  0  & 10 & 3 & 0 & 1\\
\hline\hline
\end{tabular}
\end{center}
\end{table}

\begin{figure}
\begin{centering}
\includegraphics[width=0.5\textwidth, trim=10 10 10 10, clip]{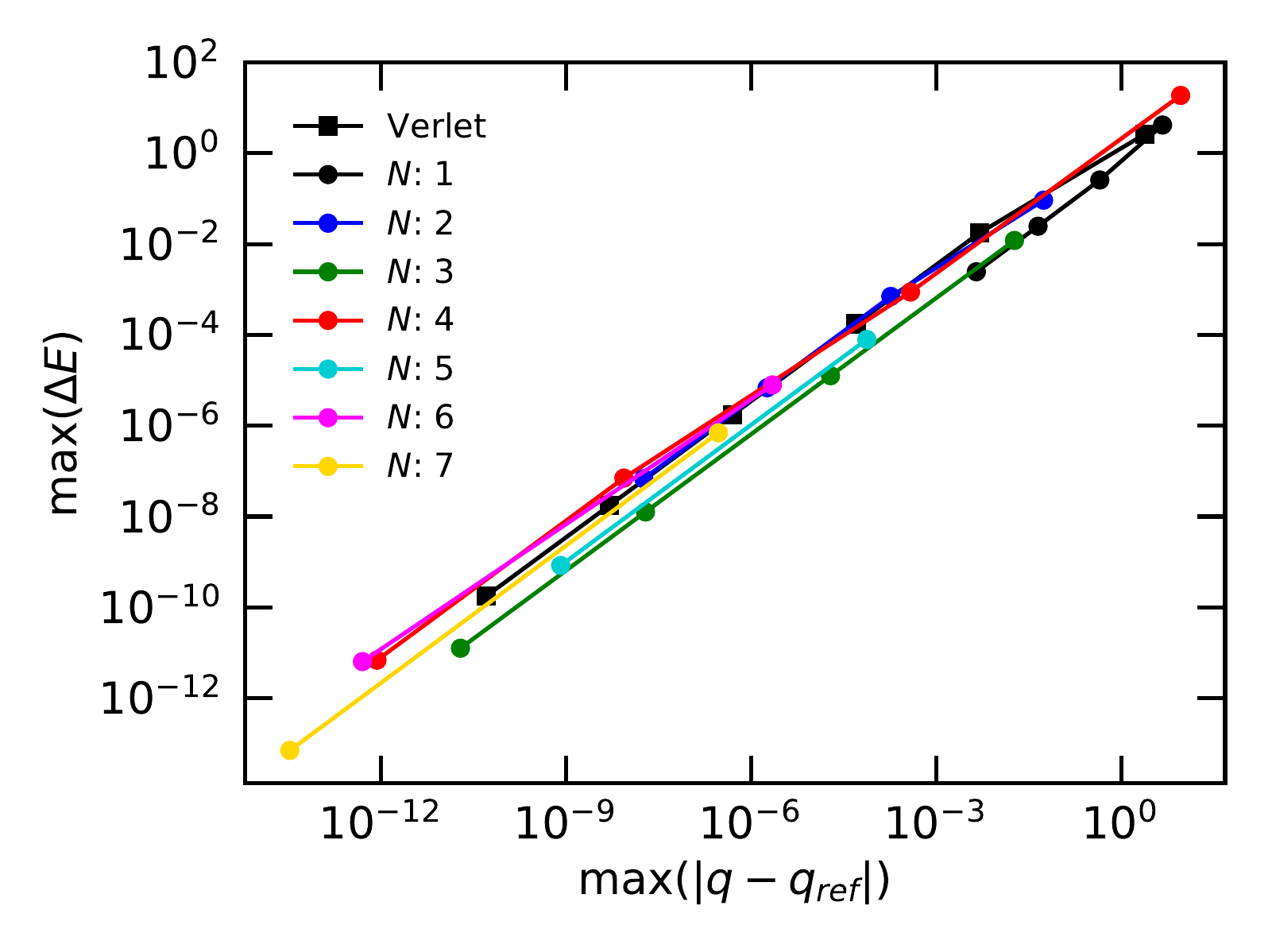}
\caption{Correlation between the maximum energy drift and the maximum RMSD for one particle in a Morse potential. The maxima are taken over the whole time grid ${t_\mathrm{0}, \dots, t_\mathrm{end}}$. The energy drift $\Delta E(t_k)$ and the RMSD are calculated using Eq.~(\ref{eq:edrift}) and the formula $|q(t_k)-q_\mathrm{ref}(t_k)|$,  respectively, where $q(t_k)$ are the computed positions in time and $q_\mathrm{ref}(t_k)$ is the analytical solution taken from Refs.~\citenum{dema1978:733, barb2007cr:543}. The measurement points (shown as circles) are connected by straight lines as guides for the eye.
\label{fig:edrift_vs_qref_q}}
\end{centering}
\end{figure}

There is an analytic solution for the classical Morse oscillator \cite{dema1978:733, barb2007cr:543} that allows  establishing a scaling relation between the maximum energy drift and the maximum root mean square deviation (RMSD) of the numerical solution from the analytical solution. As it is shown in Figure~\ref{fig:edrift_vs_qref_q} for the RMSD of the computed positions, these two accuracy measures provide equivalent views of the accuracy within a constant prefactor. 

In the following, the influence of the time step size $\Delta t$ and the spectral width parameter $\Delta L$, and the behavior of the polynomial propagators in long-time integration will be discussed.

\subsubsection{Effect of the time step}
\label{subsubsec:time-step}

\begin{figure}
\begin{centering}
\includegraphics[width=\textwidth, trim=10 10 10 10, clip]{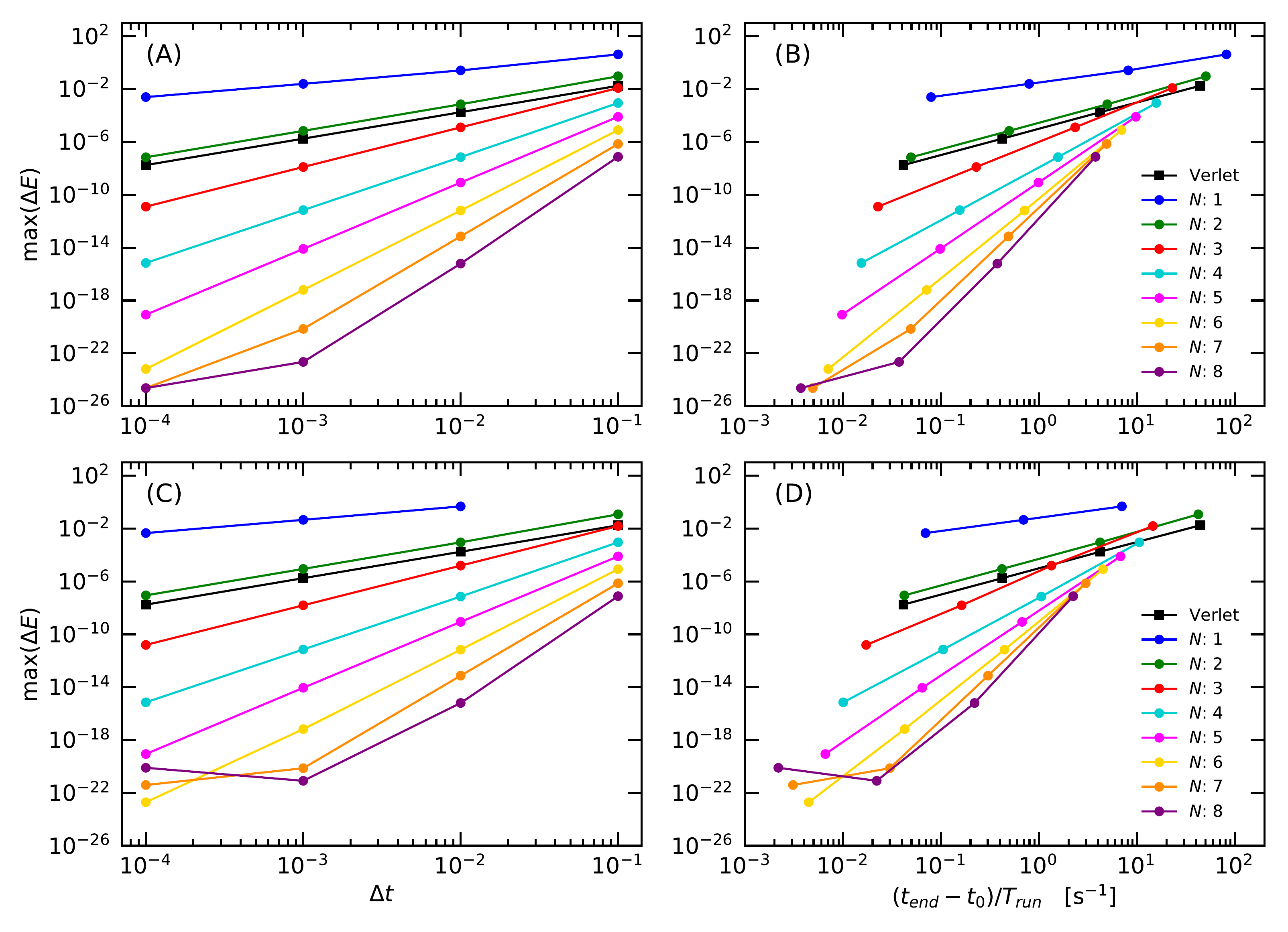}
\caption{Convergence with time step (A and C) and relation between accuracy and computational performance (B and D) for the Chebyshev (A and B) and Newton (C and D) polynomial propagators applied to one-particle Morse oscillator. The convergence and the computational performance of the velocity Verlet integrator applied to the same problem are displayed for comparison as black squares.}
\label{fig:morse-edrift-vs-tstep}
\end{centering}
\end{figure}

In Figure~\ref{fig:morse-edrift-vs-tstep},~A and C the convergence with the time step size for several expansion orders is displayed. The energy drifts with the second-order polynomial propagators decrease in a quadratic way with time step size and have therefore the same slope as the velocity Verlet integrator. The polynomial propagator of seventh order, i.e. $N=7$, yields an accuracy improvement of $14$ orders with decreasing the time step 100 times. Overall, the measured order of accuracy is determined by the expansion order, i.e the order of polynomial after which the series expansion is truncated. Compared to the Verlet integrator, the simulation with the polynomial propagators can be performed with much larger time steps at a given accuracy provided that sufficiently high number of terms in the expansion are taken (see Figure~\ref{fig:morse-edrift-vs-tstep},~A and C).

In Figure~\ref{fig:morse-edrift-vs-tstep},~A the measured energy drift for $N=7$ and $N=8$ at $\Delta t = 10^{-4}$ deviates from the systematic $\mathcal{O}((\Delta t)^N)$ convergence. This is due to round-off errors that become dominating at short time steps and high expansion orders. In double precision these round-off effects occur for energy drifts of the order of $10^{-14}$ and $10^{-12}$ with the Chebyshev and Newton propagators, respectively. To eliminate these effects, all calculations shown in Figure~\ref{fig:morse-edrift-vs-tstep} have been performed with 30 decimal digits of precision, using the \texttt{mpmath} library \cite{mpmath}. Thus, the round-off effects start limiting the accuracy of the Chebyshev propagator only as the energy drift becomes of the order of $10^{-25}$. With the Newton propagator these round-off effects (cf.~Figure~\ref{fig:morse-edrift-vs-tstep},~C) appear already for energy drifts below $10^{-23}$. With increasing the numerical precision to 60 digits (data shown in Figure~2 in the supplementary material) no round-off effects are detected within the chosen ranges of expansion orders and time step size. 

From the available results it can be concluded that the present realization of the Newton propagator is generally more sensitive to round-off effects that the Chebyshev propagator. This is also supported by experiments with other test systems, e.g. for a particle in anharmonic double-well potential and in Lennard-Jones potential (data provided in Figures~3-6 in supplementary material). Provided that there are no round-off effects, the accuracy of the two polynomial propagators is practically the same, as it is seen in Figure~\ref{fig:morse-edrift-vs-tstep},~A and C. Realistic applications of high-order polynomial propagators aim at large time steps and thus are not affected by round-off effects even in double or in single precision. Here, the high precision has been chosen to study the order of accuracy for which the time step must be decreased.

The high-order expansion will increase the computational effort per time step compared to a lower-order expansion. To assess the overall efficiency of the polynomial propagators a performance measure typically used in molecular dynamics simulations is employed, i.e. the total time of integration divided by the elapsed total computing time. This computational performance is usually measured in nanoseconds per day but here the unit s$^{-1}$ is used. Figure~\ref{fig:morse-edrift-vs-tstep},~B and D, shows the energy drift vs. computational performance. Higher-order propagators become more efficient if high accuracy is required, i.e. the high computational costs per time step are compensated by the large time steps of integration. For this test model the computational effort with the Newton propagator is up to twice larger than with the Chebyshev propagator for the same accuracy.

\subsubsection{Effect of the spectral width $\Delta L$}
\label{subsubsec:spectral-width}

\begin{figure}
\begin{centering}
\includegraphics[width=\textwidth, trim=10 10 10 10, clip]{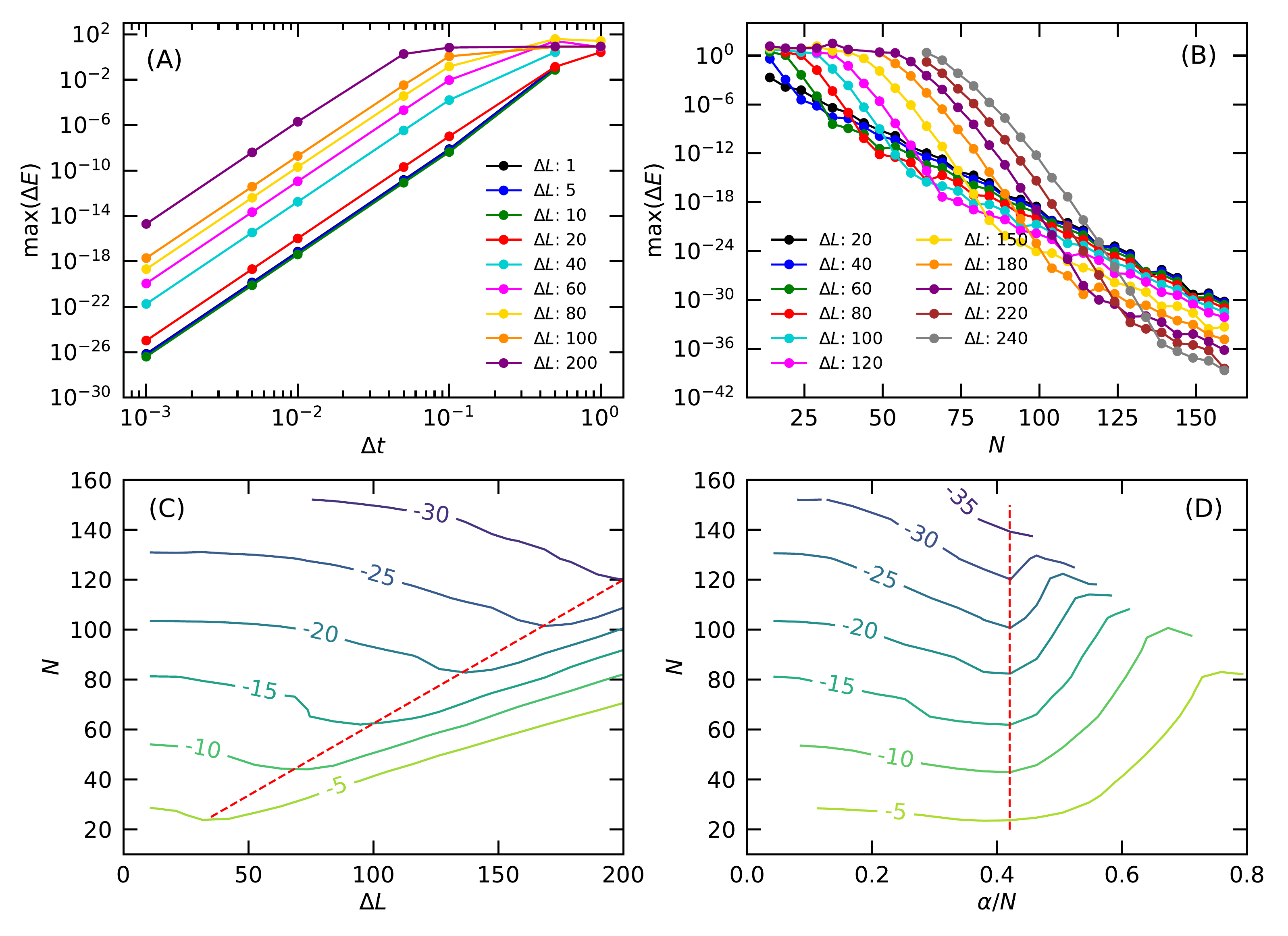}
\caption{The effect of the spectral width $\Delta L$ on the accuracy in terms of the maximum energy drift for the Chebyshev polynomial propagator applied to one-particle Morse oscillator. The number of expansion terms in plot A is fixed to $N=9$. The time step in all other plots (B, C, and D) is fixed to $\Delta t = 0.5$. The red dashed lines denote the optimal $\Delta L$ values for different $N$.}
\label{fig:delta-L}
\end{centering}
\end{figure}

While in quantum dynamics $\Delta L$ can be estimated by analyzing the matrix representation of $\mathcal{L}$ using standard numerical methods, in classical dynamics the determination of the eigenvalues of $\mathcal{L}$ is an issue \cite{zwan1966:170, kumi1984:821, carv1996a:3597, wilk1997b:27}. Moreover, the spectral bounds of $\mathcal{L}$ may even not exist in some cases, as has been claimed in previous studies \cite{zwan1966:170, spoh1975:323}. Therefore,  $\Delta L$ is treated in this study as a fit parameter and in Figure~\ref{fig:morse-edrift-vs-tstep} it is fixed arbitrarily to unity. This raises the question how $\Delta L$ can optimally be chosen \emph{a priori} in practical applications of the polynomial propagators. The $\Delta L$ parameter has indirect influence on the computational effort through the minimum number of expansion terms $N$ necessary for convergence via $N>\Delta t \Delta L/2$ (see Section~\ref{subsec:chebyshev}.

The parameter $\Delta L$ has no influence on the accuracy convergence behavior with varying $\Delta t$, as indicated by the parallel lines for different $\Delta L$ in the region of sufficiently small $\Delta t$ in Figure~\ref{fig:delta-L},~A. The increase of energy drift for larger $\Delta L$ is due to insufficient convergence with the chosen fixed expansion order $N=9$. The convergence behavior is shown in Figure~\ref{fig:delta-L},~B where the maximum energy drift is calculated for increasing expansion order. The time step $\Delta t=0.5$ is chosen sufficiently large to render the effect of $\Delta L$ and to ensure stable numerical evaluation within a precision of 60 decimal digits. This high numerical precision was also necessary to avoid numerical issues, such as round-offs and overflows, that occur for large $\Delta L$ and $N$ and mask the behavior of the numerical scheme. 

With increasing $\Delta L$ more expansion terms are needed for convergence because the Bessel function $J_n(\alpha)$ starts decreasing exponentially for  $n > \alpha$, where $\alpha = \Delta t \Delta L/2$. This behavior is already known \cite{lefo1991bc:59} but has not been studied in the case of classical dynamics, where $\Delta L$ cannot be determined \emph{a priori}. If the spectrum of $\mathcal{L}$ is bounded then the optimal choice of $\Delta L$ is the difference of the spectral bounds and larger $\Delta L$ will bring no further accuracy improvement but more expansion terms, i.e. more computational effort, will be needed for convergence. If the $\mathcal{L}$ spectrum is not bounded, then a trade-off between the selected $\Delta t$ and $\Delta L$ has to be made to ensure convergence and long-term stability with a feasible number of expansion terms.

For sufficiently large $N$, when the polynomial expansion is converged, increasing $\Delta L$ can yield better accuracy. This is the case for the Morse oscillator example shown in~Figure~\ref{fig:delta-L},~B. For $N=100$ the best accuracy is achieved with $\Delta L=150$ while the expansion with $\Delta L = 200$ is not converged, whereas for $N=120$ then $\Delta L = 200$ yields the best accuracy. The same behavior is depicted on the contour plot in Figure~\ref{fig:delta-L},~C. For example, with $N=160$ an energy drift of the order $10^{-30}$ can be realized for $\Delta L = 25$. With the same number of terms, i.e. the same computational effort, the energy drift is further reduced to  $10^{-35}$ for $\Delta L= 200$. In contrast, for $N=60$ approx. 5 orders of magnitude better accuracy is obtained with $\Delta L= 25$ than with $\Delta L= 200$.

For each number of terms there exists a different critical value for $\Delta L$ to satisfy convergence. The location of the critical value can be better analyzed by normalizing $\Delta L$ with the expansion order at fixed time step, i.e. using $\alpha/N$ which is shown in Figure~\ref{fig:delta-L}, C and D. With decreasing $\Delta L$ for any fixed $N$ the energy drift steeply decreases for $\alpha/N \gtrsim 0.5$ due to the minimal number of terms necessary to converge the expansion for given $\Delta L$. Further decrease of $\Delta L$ does not lead to lower energy drift presumably because smaller part of the $\mathcal{L}$ spectrum is included in $\Delta L$. This gives rise to an energy drift minimum occurring for $\alpha/N$ in the range between 0.4 and 0.5. The position of this minimum with increasing $N$ is marked by red dashed lines in Figure~\ref{fig:delta-L}, C and D.

Considering the results from the measurements the following two practical cases of using the polynomial propagators can be suggested: 1) With pre-selected numerical costs, i.e. $N$ and $\Delta t$ are fixed \emph{a priori}, the optimal choice of $\Delta L$ should be such that $\alpha/N$ is slightly smaller than 0.5. In this mode it is not possible to control the accuracy whereas the optimal possible accuracy at fixed costs, i.e. $\Delta t$ and $N$, can be realized. This method can be used as a coarse first-pass propagator to yield the input for a parallel-in-time integration scheme to achieve refined accuracy; 2) the expansion order $N$, the time step size $\Delta t$ and $\Delta L$ can be adapted in the course of propagation so that a predefined accuracy is achieved with applying additional conditions, such as minimum $\Delta t$ and maximum $N$, in order to restrict the optimization search.

The measurements discussed in Figure~\ref{fig:delta-L} have been performed with the Chebyshev propagator. Because the parameter $\Delta L$ enters the Newton propagator through the scaling factor, it can be assumed that it may have similar performance implications. Therefore, the recipes of choosing $\Delta L$, suggested above might be helpful as well for practical applications using the Newton propagator. Nevertheless, the Newton propagator is more sensitive to round-off errors due to limited numerical precision, which prevents the analysis of the energy drift for very large expansion orders and small values of $\Delta L$. Measurement data with the Newton propagator applied to two test models, the Morse oscillator and the anharmonic double-well potential, can be found in Figures~7-12 in the supplementary material.

\subsubsection{Long-time behavior}
\label{subsubsec:long-time}

\begin{figure}
\begin{centering}
\includegraphics[width=\textwidth, trim=10 10 10 10, clip]{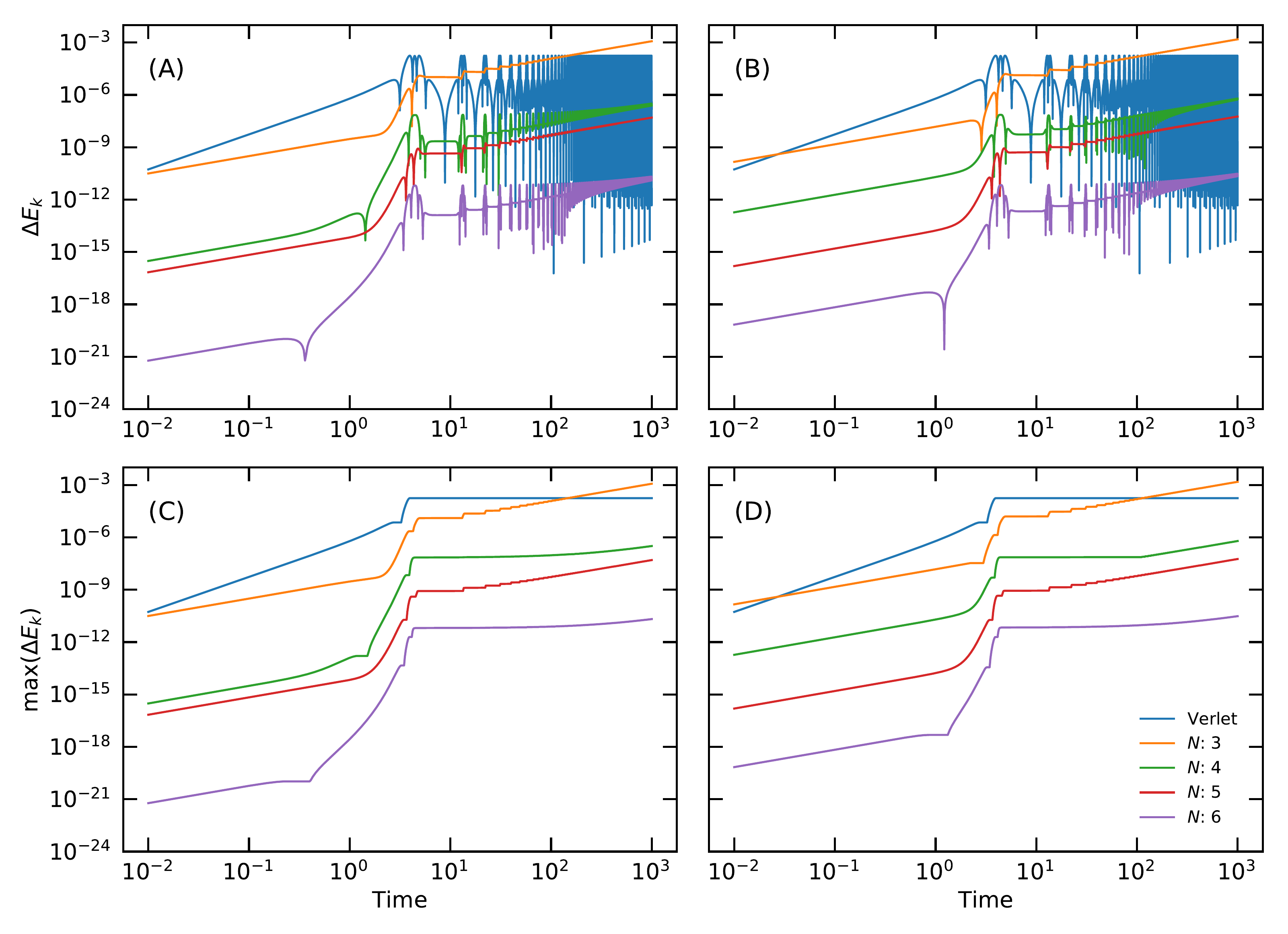}
\caption{Accuracy of the Chebyshev (A and C) and Newton (B and D) polynomial propagators measured by the energy drift (A and B) and the running maximum of the energy drift (C and D) for a long period of integration with $10^5$ time steps with size $\Delta t=0.01$. For comparison, the energy drift and the running maximum of the energy drift were measured with the velocity Verlet method with the same time step size.}
\label{fig:long-time}
\end{centering}
\end{figure}

The maximum energy drift for a relatively short, fixed period of time was used for measuring the accuracy in the discussion above. An important quality aspect of time integrators is the energy stability over long periods of integration. To study the behavior of the polynomial propagators over long time, the energy drift at every time step and the running maximum of the energy drift were plotted in Figure~\ref{fig:long-time},~A and C in the case of the Chebyshev propagator. For a comparison, these accuracy measures were plotted for the same problem integrated with the symplectic velocity Verlet method. With the first 200 time steps the energy drift with the Verlet method increases about 5 orders of magnitude from $10^{-11}$ up to $10^{-6}$. After this the energy drift starts strongly oscillating and becomes temporarily as small as $10^{-16}$, but stays constantly limited by $2\times 10^{-4}$ until the end of the simulation. 

The third-order ($N=3$) Chebyshev propagator yields a slow initial increase in the energy drift, but after $t \approx 2$ the energy drift increases rapidly and after $t \approx 4$ begins oscillating but stays limited by $1 \times 10^{-5}$ and smaller than that produced with the velocity Verlet integrator until $t \approx 12$. After this the maximum energy drift starts increasing nearly linearly with the time step and after $t \approx 150$ exceeds the energy drift of the Verlet integrator. At the end of the simulation the maximum energy drift of the third-order Chebyshev propagator is approx. $10^{-3}$.

Chebyshev propagators of higher orders $N=4, 5, 6$ have similar behavior to that with $N=3$. Particularly the time of the beginning of the plateau-like region is $t \approx 4$ for all orders. However, the overall accuracy is improved with increasing expansion order. At $t=10^3$ the Chebyshev propagators with $N\ge 4$ yield smaller maximum energy drifts than the velocity Verlet integrator. The long-term accuracy of the Newton propagator, shown in Figure~\ref{fig:long-time},~B and D, behaves similarly to the Chebyshev propagator. The accuracy improvements with increasing expansion order for short times after $t_0$ are equally spaced for the Newton propagator while the odd orders of the Chebyshev propagator yield less accuracy gain than the even orders. This is clearly identified comparing the gaps between the green and the red lines for $t < 2$ on the left and the right panels of Figure~\ref{fig:long-time}.

It should be noted that there are two striking variations in the behavior of both Chebyshev and Newton polynomial propagators for even and odd orders. Firstly, the energy drifts of even-order propagators exhibit oscillations with larger amplitudes than those of odd orders, as seen in Figure~\ref{fig:long-time},~A and B. Secondly, the propagators of even orders result in longer plateau-like regions in the maximum energy drift and thus yield more stable accuracy at long integration times (cf.~Figure~\ref{fig:long-time},~C and D). According to these observations, even-order polynomial propagators behave more similar to velocity Verlet in long-time integration.

The Chebyshev and Newton polynomial propagators exhibit very similar long-time behavior for the anharmonic double-well potential. These results are available in Figures 11 and 12 in the supplementary material.


\subsection{Many-particles system: Lennard-Jones gas}
\label{subsec:mp}

In the following, the polynomial propagators will be demonstrated on a simple example system with interacting argon atoms in Lennard-Jones potential
\begin{equation}
v(r_{jk}) = 4 \epsilon \left[\left(\frac{\sigma}{r_{jk}}\right)^{12}-\left(\frac{\sigma}{r_{jk}}\right)^6 \right]~.
\label{eq:lj-potential}
\end{equation}
The parameters $\epsilon$ and $\sigma$ in Eq.~(\ref{eq:lj-potential}), and the masses $m_j$ have been chosen for argon atoms and are unities in the Lennard-Jones units used in the simulation.

For the many-particle systems, the measured orders of accuracy are very similar to those presented above for the one-particle system and will not be discussed.
These data are provided in Figures~15-18 in the supplementary material. The most relevant questions arising from using the polynomial propagators for many-particle systems is the scaling of computation time and allocated memory with the number of particles and number of expansion terms.

\begin{figure}
\begin{centering}
\includegraphics[width=\textwidth, trim=10 10 10 10, clip]{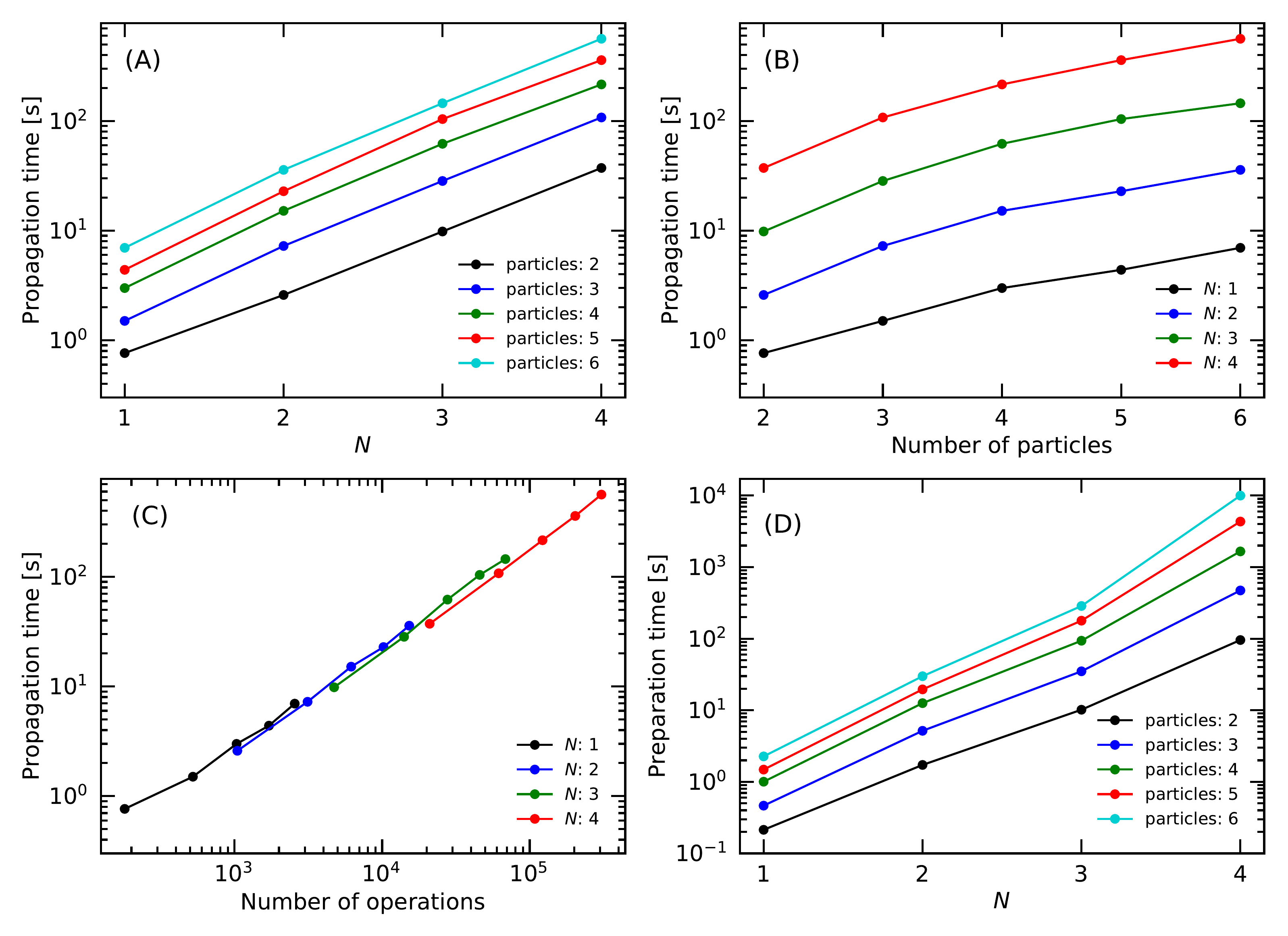}
\caption{Performance of the Chebyshev propagator in a simulation of interacting Ar atoms in Lennard-Jones potential measured for 10,000 time integration steps from $t_0=0$ to $t_\mathrm{end}=10$ using lambdifed expression (\ref{eq:mpliouville-chebyshev}) with the \texttt{NumPy} backend in double precision arithmetic. The initial conditions can be found in Table~1 in the supplementary material.}
\label{fig:many-particles}
\end{centering}
\end{figure}

Generally, the computing time for propagation scales exponentially with the number of expansion terms. While for one-particle systems the expansion is still feasible for expansion orders, as high as 160 (cf.~Figure~\ref{fig:delta-L}), this unfavorable scaling will generally limit the order of propagators used for many-particle systems. However, in Section~\ref{subsec:liouville-powers} it was shown that for a system with pair interactions the scaling should be rather $\mathcal{O}(N^6)$ in the limit of $M \gg 6$. The measured propagation times are in agreement with this theoretically estimated scaling as seen in~Figure~\ref{fig:many-particles},~A. The measured scaling is slightly improved and deviates from the exponential scaling, characterized by a straight line, with increasing the number of particles from two ($M=6$) to six ($M=18$). 

The computing time scales quadratically with the number of particles (see~Figure~\ref{fig:many-particles},~B), as expected, due to the pair interaction potentials without potential truncation. Because the focus of this study is on the time integration the interactions were evaluated ``naively'', i.e. without potential truncation. Pair interaction potentials (cf.~Eq.~(\ref{eq:pair-potential-energy})) that become constant in the limit of long distance allow potential truncation which will result in a linear scaling with the number of particles.

Figure~\ref{fig:many-particles},~C shows that the computing time correlates linearly with the number of operations determined using static analysis of the symbolic expressions before compiling. Therefore, the computing time necessary for propagation can be estimated reliably by performing analysis of the expressions and using timings and memory allocations, measured for smaller systems. Because the computing time also correlates linearly with the number of atomic objects in the expressions (not shown) the order of polynomial propagators used in practical calculations might become limited due to excessive memory allocation needed to store the expressions. The memory consumption can be reduced by partitioning of the full expressions into sub-expressions and avoid repeated storage of sub-expressions that are either identical or have the same symbolic form and only different numerical parameters. Also the kernel function compiling times will be reduced in this way.

Another important aspect of using the polynomial propagators is the preparation time which includes the computing time for i) constructing the symbolic expressions according to Eqs.~(\ref{eq:mpliouville-taylor}), (\ref{eq:mpliouville-chebyshev}) and (\ref{eq:mpliouville-newton}) and ii) compiling the high-performance kernel functions for numerical expression evaluation. In the current implementation, these two preparation stages are performed ``from scratch'' before beginning of every simulation. The preparation time shown in Figure~\ref{fig:many-particles},~D scales exponentially with the expansion order and thus can become very long for high expansion orders and many particles. Fortunately, this obvious limitation can be overcome using approaches to skip the two preparation stages. First, the expressions (\ref{eq:mpliouville-taylor}), (\ref{eq:mpliouville-chebyshev}), and (\ref{eq:mpliouville-newton}) do not depend on the specific potential energy before substitution with the explicit $A$ and $B$ expressions. The former expressions can be generated for different numbers of particles and orders and stored in libraries. On the other hand, the derivative expressions (\ref{eq:potential-energy-internal-derivs}) depend only on the order $N$ and the specific potential. These expressions can be generated for different orders and all supported force fields and stored in libraries. Upon initialization all necessary expressions can be loaded from the library and the computational costs for symbolic construction in the first preparation stage will be saved. Because the final expressions are specific for every system, the kernel functions for numerical evaluation have to be compiled after defining a new model system. However, if a simulation with the same system has to be continued from a previous run or started with different positions and momenta, the already compiled kernel functions can be reused, i.e. also the second preparation stage can be bypassed.

\section{Summary}
\label{sec:summary}

Classical numerical propagators have been devised based on expansions of the time evolution operator in series of Newton and Chebyshev polynomials. The principal advantage of the method is the possibility to increase the time step size up to several orders of magnitude with no loss of accuracy and stability. The proposed method does not require separation of time scales and is generally applicable also to problems with significant mixing of time scales because of the uniform accuracy of the proposed propagators. However, a minimum expansion order is required to achieve convergence. This minimum expansion order is proportional to the time step size $\Delta t$ and the spectral width $\Delta L$ of the Liouville operator. This gives rise to the main drawback of the method: High expansion orders are necessary in order to increase the time step for systems with large spectral width. After this convergence threshold, the accuracy converges exponentially with the number of terms included in the expansion after a certain number of leading terms, necessary for convergence determined by the time step size $\Delta t$ and the spectral width $\Delta L$. Higher orders of accuracy beyond the convergence threshold are usually not necessary for molecular dynamics simulations based on approximate empirical force fields.

The method has been applied to a test model with one particle in Morse potential and many particles with Lennard-Jones interactions. In the numerical experiments it has been shown that the polynomial propagators have orders of accuracy identical with the highest polynomial order in the truncated expansion, i.e. $\mathcal{O}((\Delta t)^N)$ whereby the Newton and Chebyshev propagators exhibit the same accuracy convergence. Additionally it has been shown that the polynomial propagators exhibit long-time stability for the test models already for low expansion orders, e.g. for $N=4$.

The influence of the spectral width $\Delta L$, which is varied arbitrarily in this work, on the accuracy has been investigated. Large eigenvalues of the Liouville operator may not be included in $\Delta L$ and thus limit the accuracy and long-time stability. A practical recipe has been suggested to choose this parameter properly. The proper bound of this parameter is essential for the stability of the algorithm, and, when the parameter is too large, for the applicability of the method. Therefore, an extended theoretical study of propagator's properties related to this parameter and to the spectrum of the Liouville operator for classical systems is needed.

The proposed method is suitable for integrating the equations of motion of any model system described by classical force fields with potential energy in analytic form. Thereby, an issue arises from the rapidly increasing number of mixed partial derivatives when the method is applied to many-particle systems. It was shown that in the case of pair-interaction potentials the mixing in high-order derivatives is limited to only six degrees of freedom reducing the scaling with the expansion order $N$ to only $\mathcal{O}(N^6)$ for sufficiently large number of particles, i.e. for $M \gg 6$. In future work, the scheme can be extended to treat bonded interactions to fully adopt the classical force fields used in molecular dynamics.

\section{Outlook}
\label{sec:outlook}

Besides increased accuracy and time step size, the polynomial propagators provide the possibility to perform the numerical evaluations within every time step in parallel and thus to reduce the time to solution. With the velocity Verlet integrator this is also possible but the computational workload that can effectively be parallelized is the evaluation of the forces $-A^{(1)}_j(q_1, \dots, q_M)$ which is obviously much smaller than that of the polynomial propagators in Eq.~(\ref{eq:mpliouville-taylor}), (\ref{eq:mpliouville-chebyshev}) and (\ref{eq:mpliouville-newton}). The high number of numerical evaluations of the expression terms can be readily performed asynchronously on computing accelerators such as GPUs by compiling the expression evaluation kernels using generic packages such as \texttt{Theano} and \texttt{TensorFlow}. Although the parallelization of the polynomial propagators is beyond the scope of this work it is clearly a benefit that should be exploited to speed up molecular dynamics on high performance computing clusters.

Properties of the propagators that have not been discussed in this paper are the time reversal symmetry, $U(t) U(-t)=1$, and the related unitarity, $U(t) U^\dagger(t)=1$ where $U$ is the propagator approximating $e^{i\mathcal{L} t}$. Time reversal symmetry has been shown to improve long-term stability in terms of energy conservation \cite{mcla1995:151, hut1995mm:l93, funa1996hm:112, hut1997fk:26, mcla1998qt:586, hair2005s:1838}. It can be shown (see Appendix~\ref{appendix-symmetry}) that the infinite-order Chebyshev propagator is unitary and time-reversible. However, in practice the series is truncated after some order $N$ and therefore the time reversal symmetry and unitarity are not preserved. Furthermore, the Newtonian propagator does not have time reversal symmetry even in the limit of infinite expansion order. The lack of time reversal symmetry is the reason for the energy drifts observed in the long-time runs for both propagators. To achieve time reversal symmetry, midpoint or central difference time-symmetric two-step propagators can be constructed using Chebyshev polynomials of even and odd parity, respectively. Alternatively, an implicit self-starting (one-step) symmetric propagator can be constructed by following the strategy of Hut et al. \cite{hut1997fk:26}.

The current study is restricted to the microcanonical ensemble for which the spectrum of the Liouville operator is real.  However, the polynomial propagator approach is not limited to this specific case. In the presence of a thermostat, the spectrum of the Liouville operator is complex. The Newton polynomial propagator can be employed in this more general case without modifications, however, the set of interpolation points (on the complex plane) has to be constructed with care. Furthermore, Chebyshev polynomials can be replaced by Faber polynomials that can be regarded as a generalization of the Chebyshev polynomials on the complex plane. Faber polynomials have been successfully used in dissipative quantum dynamics simulations \cite{huis1998pk:29, huis1999pk:5538, xie2000cg:5263}.

\section*{Acknowledgements}

The author acknowledges support by the state of Baden-W\"urttemberg through bwHPC.

\appendix

\section{Approximating $e^{iax}$ using Chebyshev polynomials}
\label{appendix-chebyshev}

\subsection{Identities of Bessel functions and Chebyshev polynomials}

The following identities will be used. The Bessel functions of the first kind $J_n(a)$ and the modified Bessel functions of the first kind $I_n(a)$ are related by the following identities:

\begin{equation}
I_n(ia) = i^n J_n(a), \quad J_n(a) = (-i)^n I_n(ia)
\end{equation}
and
\begin{equation}
J_n(ia) = i^n I_n(a), \quad I_n(a) = (-i)^n J_n(ia),
\end{equation}
where $a \in \mathbb{R}$.

The Chebyshev polynomials of the first kind $\Psi_n(x)$ are defined for $x \in [-1, 1]$ and have the recurrence relation
\begin{equation}
\Psi_n(x) = 2x \Psi_{n-1}(x) - \Psi_{n-2}(x)
\label{eq:recu-Psi-nx}
\end{equation}
with recursion starting with $\Psi_0(x) = 1$ and $\Psi_1(x) = x$. The Chebyshev polynomials are orthogonal:
\begin{equation}
\int\limits_{-1}^1 \frac{\Psi_n(x) \Psi_m(x)}{\sqrt{1-x^2}}dx = \delta_{nm}(1+\delta_{n0})\frac{\pi}{2}
\label{eq:ortho-Tnx}
\end{equation}

\subsection{Derivation of the expansion coefficients}

The exponential function $e^{iax}$ can be expanded in two different ways 
\begin{equation}
e^{iax} = \sum\limits_{n=0}^\infty c_n(ia) \Psi_n(x)
\label{eq:expansion-Psi-nx}
\end{equation}

\begin{equation}
e^{iax} = \sum\limits_{n=0}^\infty c'_n(a) \Psi_n(ix)
\label{eq:expansion-Psi-nix}
\end{equation}
While the two expansions are equivalent and the expansion coefficients will be of the same form, the expansion (\ref{eq:expansion-Psi-nx}) is more suitable for determination of the coefficients $c_n$ whereas expansion (\ref{eq:expansion-Psi-nix}) is more suitable to expand the time evolution operator (the exponential propagator) by applying the operator $i\mathcal{L}$ using the recurrence relation (\ref{eq:recu-Psi-nx}).

The coefficients in Eq.~(\ref{eq:expansion-Psi-nx}) can be determined when it is multiplied on both sides with $\Psi_m(x)/\sqrt{1-x^2}$ and integrated over the interval $[-1, 1]$:

\begin{equation}
\int\limits_{-1}^1 \frac{e^{iax} \Psi_m(x)}{\sqrt{1-x^2}} dx = \sum\limits_{n=0}^\infty c_n(ia) \int\limits_{-1}^1 \frac{\Psi_n(x)\Psi_m(x)}{\sqrt{1-x^2}} dx.
\label{eq:expansion-Psi-nx-1}
\end{equation}
The integral on left hand side yields $\pi i^m J_m(a)$ and the one on the right hand side yields $\delta_{nm}(1+\delta_{m0})\pi/2$ so that
\begin{equation}
\pi i^m J_m(a) = c_m(ia) (1+\delta_{m0})\pi/2.
\label{eq:expansion-Psi-nx-2}
\end{equation}
The thus obtained coefficients $c_n(ia) = (2-\delta_{n0}) i^n J_n(a)$ are inserted in Eq.~(\ref{eq:expansion-Psi-nx}) resulting in
\begin{equation}
e^{iax} = \sum\limits_{n=0}^\infty (2-\delta_{n0}) i^n J_n(a) \Psi_n(x).
\label{eq:expansion-Psi-nx-3}
\end{equation}
The modified Chebyshev polynomials $\Phi(ix)$ are defined as $\Phi_n(ix) = i^n \Psi_n(x)$ \cite{tale1986:11}. The recurrence relation for these polynomials can be derived in a straightforward way from the definition
\begin{equation}
\Phi_n(ix) = 2ix \Phi_{n-1}(ix) + \Phi_{n-2}(ix)
\label{eq:recu-phinix}
\end{equation}
and the recursion is started with $\Phi_0(ix)=1$ and $\Phi_1(ix)=ix$. Using the polynomials $\Phi_n(ix)$ Eq.~(\ref{eq:expansion-Psi-nx-3}) takes the form
\begin{equation}
e^{iax} = \sum\limits_{n=0}^\infty (2-\delta_{n0}) J_n(a) \Phi_n(ix).
\label{eq:expansion-phinix}
\end{equation}
Similarly, it can be shown that in Eq.~(\ref{eq:expansion-Psi-nix}) the coefficients $c'_n(a) = I_n(a)$, i.e. 
\begin{equation}
e^{iax} = \sum\limits_{n=0}^\infty (2-\delta_{n0}) I_n(a) \Psi_n(ix).
\label{eq:expansion-Psi-nix-1}
\end{equation}
Therefore, Eq.~(\ref{eq:expansion-Psi-nix-1}) can be used alternatively to Eq.~(\ref{eq:expansion-phinix}) to approximate $e^{iax}$.

\section{Derivation of $(i\mathcal{L})^n$ in the case of one particle}
\label{appendix-one-dimension}

The evaluation of the monomials $(i\mathcal{L})^n$ for the case of one particle with position $q$ and momentum $p$ will be shown as an example. Assuming $H(q,p) = V(q)+T(p)$ and using the notation
\begin{equation}
a_k=\frac{\partial^k}{\partial q^k},\quad
b_k=\frac{\partial^k}{\partial p^k},\quad
A_k=\frac{\partial^k V}{\partial q^k},\quad
B_k=\frac{\partial^k T}{\partial p^k}
\end{equation}
equation (\ref{eq:poisson}) is transformed to
\begin{equation}
i\mathcal{L} z = (B_1 a_1 - A_1 b_1) z.
\label{eq:liouville1}
\end{equation}
By using the properties for the differential operators, and $a_k a_l=a_{k+l}$, $b_k b_l=b_{k+l}$, $a_k A_l = A_{k+l}$, $b_k B_l = B_{k+l}$, $a_k B_l = 0$ and $b_k A_l = 0$, the first powers of $i\mathcal{L}$ can be derived by successive applications of $i\mathcal{L}$, i.e. $(i\mathcal{L})^n z = i\mathcal{L}(i\mathcal{L})^{n-1} z$. Assuming that $z(q, p) = z_q(q) + z_p(p)$, so that all terms with $a_k b_l z$ cancel, the following expressions are obtained:
\begin{eqnarray}
(i\mathcal{L})^2 z & = &
B_1^2 a_2 z + A_1^2 b_2 z
- A_1 B_2 a_1 z - A_2 B_1 b_1 z, \\
\label{eq:liouville2}
(i\mathcal{L})^3 z & = &
B_1^3 a_3 z - A_1^3 b_3 z - 3 A_1 B_1 B_2 a_2 z + 3 A_1 A_2 B_1 b_2 z \nonumber \\
&& + (A_1^2 B_3 - A_2 B_1 B_2) a_1 z
+ (A_1 A_2 B_2 - A_3 B_1^2) b_1 z, \\
\label{eq:liouville3}
(i\mathcal{L})^4 z & = &
B_1^4 a_4 z + A_1^4 b_4 z - 6 A_1 B_1^2 B_2 a_3 z - 6 A_1^2 A_2 B_1 b_3 z  \nonumber \\
&& + (4 A_1^2 B_1 B_3 - 4 A_2 B_1^2 B_2 + 3 A_1^2 B_2^2) a_2 z
+ (4 A_ 1 A_3 B_1^2 + 3 A_2^2 B_1^2 - 4 A_1^2 A_2 B_2) b_2 z \nonumber \\
&& + (3 A_1 A_2 B_1 B_3 - A_3 B_1^2 B_2 - A_1^3 B_4 + A_1 A_2 B_2^2) a_1 z \nonumber \\
&& + (3 A_1 A_3 B_1 B_2 - A_1^2 A_2 B_3 - A_4 B_1^3 + A_2^2 B_1 B_2) b_1 z.
\label{eq:liouville4}
\end{eqnarray}
Equations~(\ref{eq:liouville1}-\ref{eq:liouville4}) can be generalized by induction in the following:
\begin{eqnarray}
(i\mathcal{L})^n_a z = & &  \sum_{l=1}^n \sum_{\{\alpha, \beta\}_l} F_{l, \{\alpha, \beta\}_l} \prod_{k=1}^n A_k^{\alpha_{kl}} B_k^{\beta_{kl}} a_l z 
\label{eq:ilna} \\
(i\mathcal{L})^n_b z  = & (-1)^n & \sum_{l=1}^n \sum_{\{\alpha, \beta\}_l} F_{l, \{\alpha, \beta\}_l} \prod_{k=1}^n B_k^{\alpha_{kl}} A_k^{\beta_{kl}} b_l z
\label{eq:ilnb}
\end{eqnarray}
where $(i\mathcal{L})^n_a z + (i\mathcal{L})^n_b z = (i\mathcal{L})^n z$ and the powers $\alpha_{kl}$ and $\beta_{kl}$ are non-negative solutions of the Diophantine equations
\begin{equation}
\sum_{k=1}^n k\alpha_{kl} + l = n;\quad 
\sum_{k=1}^n k\beta_{kl} = n; \quad
\sum_{k=1}^n(\alpha_{kl}+\beta_{kl}) = n
\label{eq:diophantine}
\end{equation}
for every term of order $l$ and the summation index $\{\alpha, \beta\}_l$ denotes the set of all non-negative integer solutions of Eq.~(\ref{eq:diophantine}).

The expressions (\ref{eq:ilna}) and (\ref{eq:ilnb}) can be simplified when $(i\mathcal{L})^n$ is applied to $q$ and to $p$ because only the terms for $l=1$ are remaining:
\begin{equation}
(i\mathcal{L})^n q = \sum_{\{\alpha, \beta\}} F_{\{\alpha, \beta\}} \prod_{k=1}^n A_k^{\alpha_k} B_k^{\beta_{k}}, \quad
(i\mathcal{L})^n p = (-1)^n \sum_{\{\alpha, \beta\}} F_{\{\alpha, \beta\}} \prod_{k=1}^n B_k^{\alpha_{k}} A_k^{\beta_{k}}
\label{eq:ilnqp}
\end{equation}
Finally, the expressions for $(i\mathcal{L})^n q$ and $(i\mathcal{L})^n p$ in Eq.~(\ref{eq:ilnqp}) are symbolically equivalent, i.e. $(i\mathcal{L})^n p$ can be obtained from $(i\mathcal{L})^n q$ with replacing all $B$ symbols with $A$ symbols, and $A$ symbols with $B$ symbols. After that, futher simplifications can be performed taking into account that $B_k = 0$ for $k>2$ if $T(p) = p^2/(2m)$ with $m$ being the mass of the particle.

\section{Time reversal symmetry and unitarity}
\label{appendix-symmetry}

Considering the truncated series $U_N(t) = \sum\limits_{n=0}^N   (2-\delta_{n0}) i^n J_n(a)\Psi_n(x)$ the identity $U_N^\dagger(t) = U_N(-t)$ is fulfilled for any $N$:
\begin{eqnarray}
U_N^\dagger(t) & = & \left[\sum\limits_{n=0}^N (2-\delta_{n0}) J_n(t) i^n \Psi_n(\mathcal{L})\right]^\dagger 
                               =   \sum\limits_{n=0}^N (2-\delta_{n0}) J_n(t) (-i)^n \Psi_n(\mathcal{L}) \\
                             & = & \sum\limits_{n=0}^N (2-\delta_{n0}) (J_n(t)(-1)^n) i^n \Psi_n(\mathcal{L})
                               =  \sum\limits_{n=0}^N (2-\delta_{n0}) J_n(-t) i^n \Psi_n(\mathcal{L}) \\
                             & = & U_N(-t)
\end{eqnarray}
Here it is used that $\mathcal{L}$ is Hermitian, i.e. $\mathcal{L}^\dagger = \mathcal{L}$. It will be shown now that $U_\infty(t) U_\infty(-t) = 1$. After replacing $\mathcal{L}\rightarrow \lambda$ and integrating both sides, one gets:
\begin{eqnarray}
\sum\limits_{m,n=0}^\infty (2-\delta_{n0}) J_n(t) i^n \Psi_n(\lambda) (2-\delta_{m0}) J_m(-t) i^m \Psi_m(\lambda) & = & 1 \\
\sum\limits_{m,n=0}^\infty (2-\delta_{n0}) (2-\delta_{m0}) i^{m+n} (-1)^m J_m(t) J_n(t) \int\limits_{-1}^1 \frac{\Psi_m(\lambda) \Psi_n(\lambda)}{\sqrt{1-\lambda^2}} d\lambda & = & \int\limits_{-1}^1 \frac{1}{\sqrt{1-\lambda^2}} d\lambda \\
\sum\limits_{n=0}^\infty (2-\delta_{n0})^2 i^{2n} (-1)^n J^2_n(t) (1+\delta_{n0})\frac{\pi}{2} & = & \pi \\
\sum\limits_{n=0}^\infty (2-\delta_{n0}) J^2_n(t) & = & 1 \label{besselj-ident}
\end{eqnarray}
Eq.~(\ref{besselj-ident}) is an identity for the Bessel functions which is exact only for the infinite series. From this it follows that $U_\infty(t) U_\infty^\dagger(t)=1$, i.e. the Chebyshev polynomial propagator is unitary for $N\rightarrow\infty$. Because unitarity implies $U_\infty^\dagger(t) = U_\infty^{-1}(t)$ the Chebyshev polynomial propagator is time-symmetric according to the definition $U^{-1}_\infty(t) = U_\infty(-t)$ \cite{hair2005s:1838, mcla1995:151, mcla1998qt:586}.



\end{document}


\title{Supplimentary material for the manuscript \\ ``Polynomial propagators for classical molecular dynamics''}
\author{Ivan Kondov}
\affiliation{Scientific Computing Center, Karlsruhe Institute of Technology, Hermann-von-Helmholtz-Platz 1,
76344 Eggenstein-Leopoldshafen, Germany}
\date{\today}

\maketitle

\newpage

\begin{turnpage}
\begin{table}[ht]
\caption{
  Summary of the system and simulation parameters for the different systems used in the measurements.
  \label{tab:params-summary}}
\begin{ruledtabular}
\begin{tabular}{llrrrrr}
System               & Potential energy & $m_i$ & $t_0$ & $t_\mathrm{end}$ & Initial positions & Initial momenta  \\
\hline
Morse  1p            & $V(q) = e^{-2 (q-1)} - 2 e^{- (q-1)}$  &1 & 0 & 10 & 3 & 0\\
Anharmonic 1p     & $V(q) = \frac{1}{2} q^4 - q^2$ &1 & 0 & 10 & 0 & 0.1 \\
Lennard-Jones 1p & $V(q) = 4 [(1/q)^{12}-(1/q)^6]$ &1 & 0 & 2 & 2 & -1 \\ \hline
Morse 2p              & $V(r_{12}) = e^{-2 (r_{12}-1)} - 2 e^{- (r_{12}-1)}$ &2 & 0 & 10 & (1, 0, 0), (4, 0, 0) & (0, 0, 0), (0, 0, 0) \\
Lennard-Jones 2p  & $V(r_{12}) = 4 [(1/r_{12})^{12}-(1/r_{12})^6]$ &1 & 0 & 10 & (1, 2, 3), (3, 2, 3) & (0, 0, 0), (-1, 0, 0) \\ \hline
\multirow{2}{*}{Lennard-Jones mp} & \multirow{2}{*}{$V = \sum_{i<j} 4 [(1/r_{ij})^{12}-(1/r_{ij})^6]$
\footnote{The distances $r_{ij}$ are defined as $r_{ij} = \sqrt{(q_{ix}-q_{jx})^2+(q_{iy}-q_{jy})^2+(q_{iz}-q_{jz})^2}$.}}
 & \multirow{2}{*}{1} & \multirow{2}{*}{0} & \multirow{2}{*}{10} & \multirow{2}{*}{
 \shortstack[l]{(0, 0, 0), (2, 0, 0), (0, 2, 0),\\ (0, 0, 2), (2, 2, 0), (0, 2, 2)}} 
 & 
 \multirow{2}{*}{\shortstack[l]{(0, 0, 0), (-1, 0, 0), (0, -1, 0), \\  (0, 0, -1), (-1, -1, 0), (0, -1, -1)}} \\
                            &  &  &   &  &  & \\
\end{tabular}
\end{ruledtabular}
\end{table}
\end{turnpage}

\begin{table}
\caption{Overview of the figures in the supplementary information. \label{tab:figures-summary}}
\begin{ruledtabular}
\begin{tabular}{rlllll}
Figure  & System (cf.~Tab.~\ref{tab:params-summary}) & Propagator & Precision\footnote{In decimal digits. 16 digits corresponds to standard double precision.} & Tool\footnote{For numerical expression evaluation, \mbox{see https://docs.sympy.org/latest/modules/numeric-computation.html}} & Backend\footnote{For numerical expression evaluation, \mbox{see https://docs.sympy.org/latest/modules/numeric-computation.html}} \\ \hline
\ref{fig:morse-edrift-vs-tstep-prec=16} & Morse 1p & CHEB/NEWT & 16 & \texttt{subs/evalf} &  \texttt{math} \\
\ref{fig:morse-edrift-vs-tstep-prec=60} & Morse 1p & CHEB/NEWT & 60 & \texttt{subs/evalf} &  \texttt{mpmath} \\
\ref{fig:anharmonic-edrift-vs-tstep-prec=30}  & Anharmonic 1p & CHEB/NEWT & 30 & \texttt{subs/evalf} & \texttt{mpmath} \\
\ref{fig:anharmonic-edrift-vs-tstep-prec=60}  & Anharmonic 1p & CHEB/NEWT & 60 & \texttt{subs/evalf} & \texttt{mpmath} \\
\ref{fig:lj-edrift-vs-tstep-prec=30}  & Lennard-Jones 1p &  CHEB/NEWT  & 30 & \texttt{subs/evalf} & \texttt{mpmath} \\
\ref{fig:lj-edrift-vs-tstep-prec=60}  & Lennard-Jones 1p &  CHEB/NEWT  & 60 & \texttt{subs/evalf} & \texttt{mpmath} \\
\ref{fig:delta-L-newton-morse-prec=60}  & Morse 1p & NEWT & 60 & \texttt{subs/evalf} & \texttt{mpmath} \\
\ref{fig:delta-L-newton-morse-prec=90}  & Morse 1p & NEWT & 90 & \texttt{subs/evalf} & \texttt{mpmath} \\
\ref{fig:delta-L-chebyshev-anharmonic-prec=30}  & Anharmonic 1p & CHEB & 30 & \texttt{subs/evalf} & \texttt{mpmath} \\
\ref{fig:delta-L-chebyshev-anharmonic-prec=60}  & Anharmonic 1p & CHEB & 60 & \texttt{subs/evalf} & \texttt{mpmath} \\
\ref{fig:delta-L-newton-anharmonic-prec=30}  & Anharmonic 1p & NEWT & 30 & \texttt{subs/evalf} & \texttt{mpmath} \\
\ref{fig:delta-L-newton-anharmonic-prec=60}  & Anharmonic 1p & NEWT & 60 & \texttt{subs/evalf} & \texttt{mpmath} \\
\ref{fig:long-time-anharmonic-dp}  & Anharmonic 1p & CHEB/NEWT & 16 & \texttt{subs/evalf} & \texttt{math} \\
\ref{fig:long-time-anharmonic-prec=30}  & Anharmonic 1p & CHEB/NEWT & 30 & \texttt{subs/evalf} & \texttt{mpmath} \\ \hline
\ref{fig:2p-morse-edrift-vs-tstep-prec=30}  & Morse 2p & CHEB/NEWT & 30 & \texttt{lambdify} & \texttt{mpmath} \\
\ref{fig:2p-lj-edrift-vs-tstep-prec=30}  & Lennard-Jones 2p & CHEB/NEWT & 30 & \texttt{lambdify} & \texttt{mpmath} \\
\ref{fig:2p-lj-edrift-vs-tstep-dp-lambdify}  & Lennard-Jones 2p & CHEB/NEWT & 16 & \texttt{lambdify} & \texttt{numpy} \\
\ref{fig:2p-lj-edrift-vs-tstep-dp-theano}  & Lennard-Jones 2p & CHEB/NEWT & 16 & \texttt{theano} & \texttt{theano} \\
\ref{fig:many-particles-newton}  &  Lennard-Jones mp & NEWT & 16 & \texttt{lambdify} & \texttt{numpy} \\
\end{tabular}
\end{ruledtabular}
\vspace{1ex}
\end{table}

\pagebreak

\begin{figure}
\includegraphics[width=\textwidth, trim=10 10 10 10, clip]{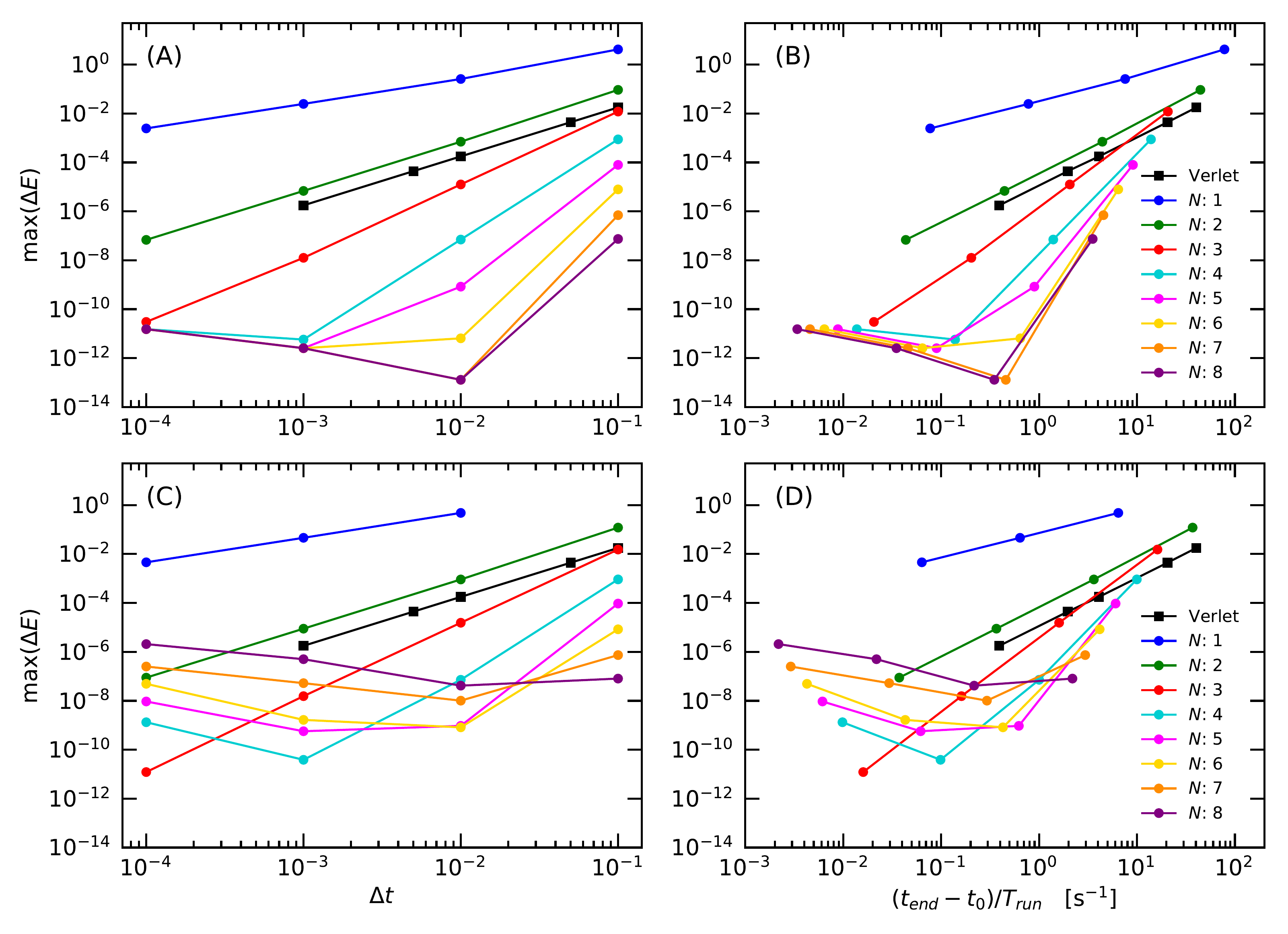}
\caption{Convergence with time step (A, C) and relation between accuracy and computational efficiency (B, D) for the Chebyshev (A, B) and Newton (C, D) polynomial propagators applied to the Morse oscillator. The  convergence and the computational efficiency of the velocity Verlet integrator applied to the same problem are shown for comparison as black squares. The expressions were evaluated in standard double precision.}
\label{fig:morse-edrift-vs-tstep-prec=16}
\end{figure}

\begin{figure}
\includegraphics[width=\textwidth, trim=10 10 10 10, clip]{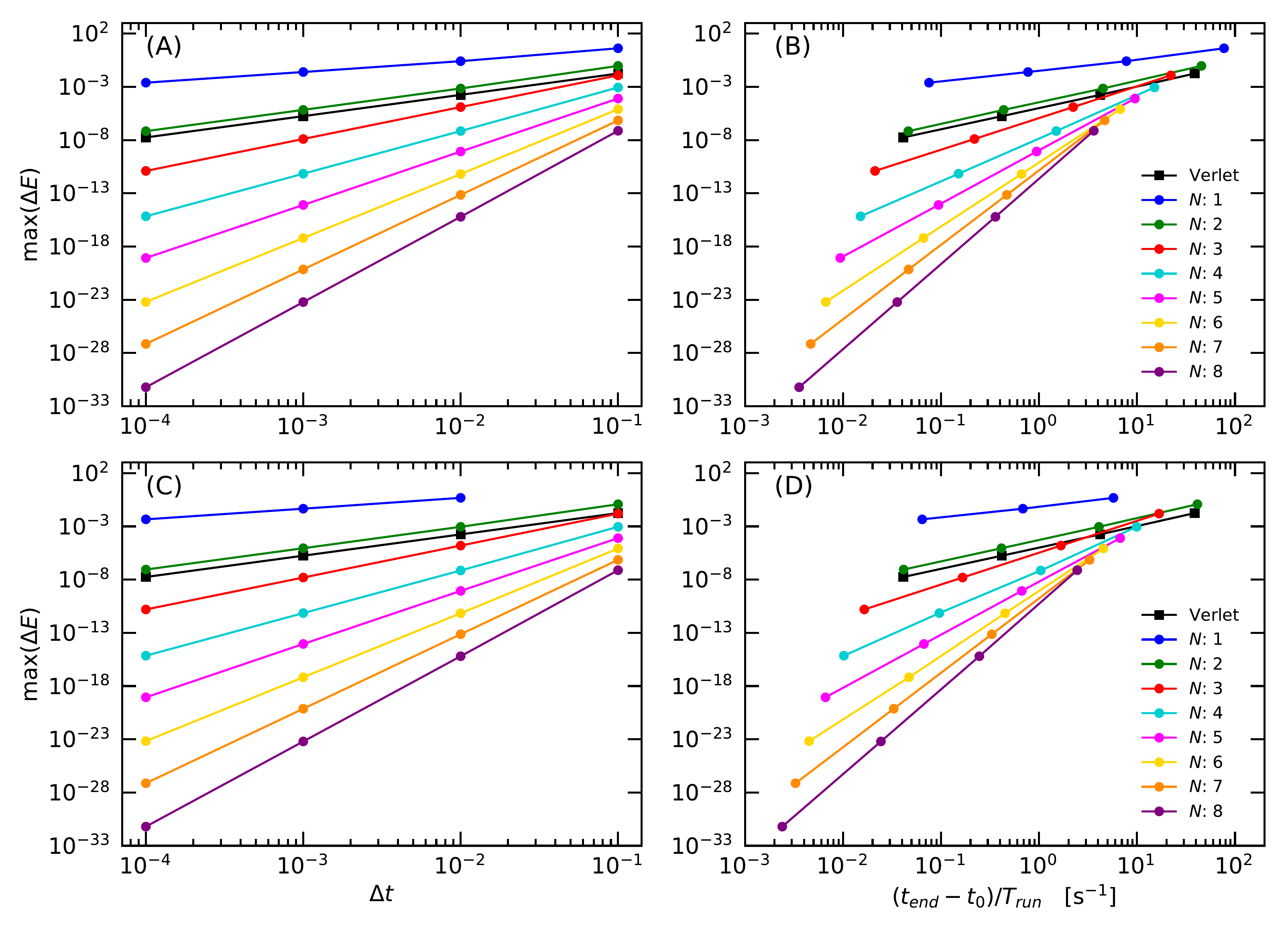}
\caption{Convergence with time step (A, C) and relation between accuracy and computational efficiency (B, D) for the Chebyshev (A, B) and Newton (C, D) polynomial propagators applied to the Morse oscillator. The  convergence and the computational efficiency of the velocity Verlet integrator applied to the same problem are shown for comparison as black squares. The expressions were evaluated in precision of 60 digits using \texttt{mpmath}.}
\label{fig:morse-edrift-vs-tstep-prec=60}
\end{figure}

\begin{figure}
\includegraphics[width=\textwidth, trim=10 10 10 10, clip]{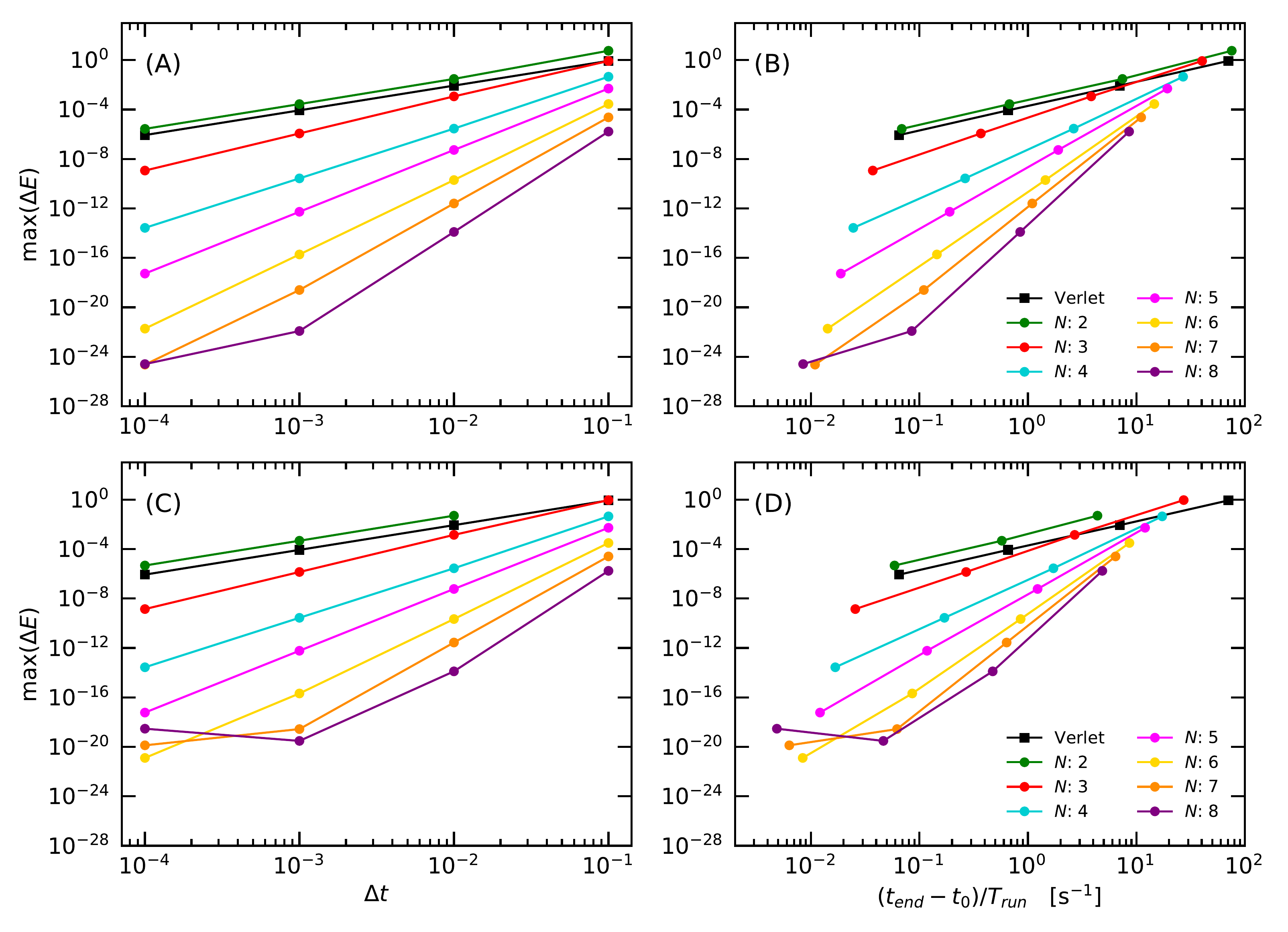}
\caption{Convergence with time step (A, C) and relation between accuracy and computational efficiency (B, D) for the Chebyshev (A, B) and Newton (C, D) polynomial propagators applied to the anharmonic oscillator. The  convergence and the computational efficiency of the velocity Verlet integrator applied to the same problem are shown for comparison as black squares. The expressions were evaluated in precision of 30 digits using \texttt{mpmath}.}
\label{fig:anharmonic-edrift-vs-tstep-prec=30}
\end{figure}

\begin{figure}
\includegraphics[width=\textwidth, trim=10 10 10 10, clip]{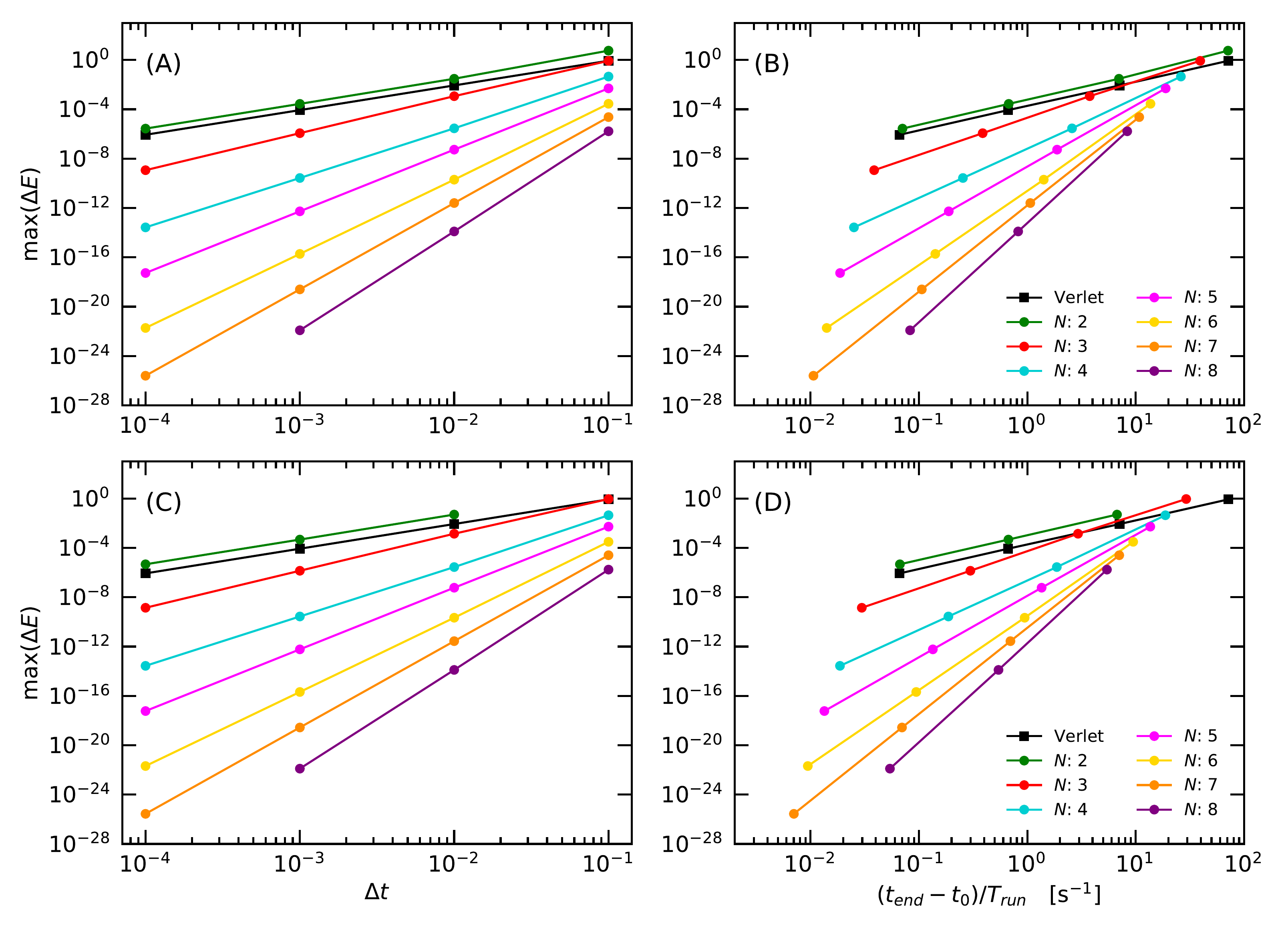}
\caption{Convergence with time step (A, C) and relation between accuracy and computational efficiency (B, D) for the Chebyshev (A, B) and Newton (C, D) polynomial propagators applied to the anharmonic oscillator. The  convergence and the computational efficiency of the velocity Verlet integrator applied to the same problem are shown for comparison as black squares. The expressions were evaluated in precision of 60 digits using \texttt{mpmath}.}
\label{fig:anharmonic-edrift-vs-tstep-prec=60}
\end{figure}

\begin{figure}
\includegraphics[width=\textwidth, trim=10 10 10 10, clip]{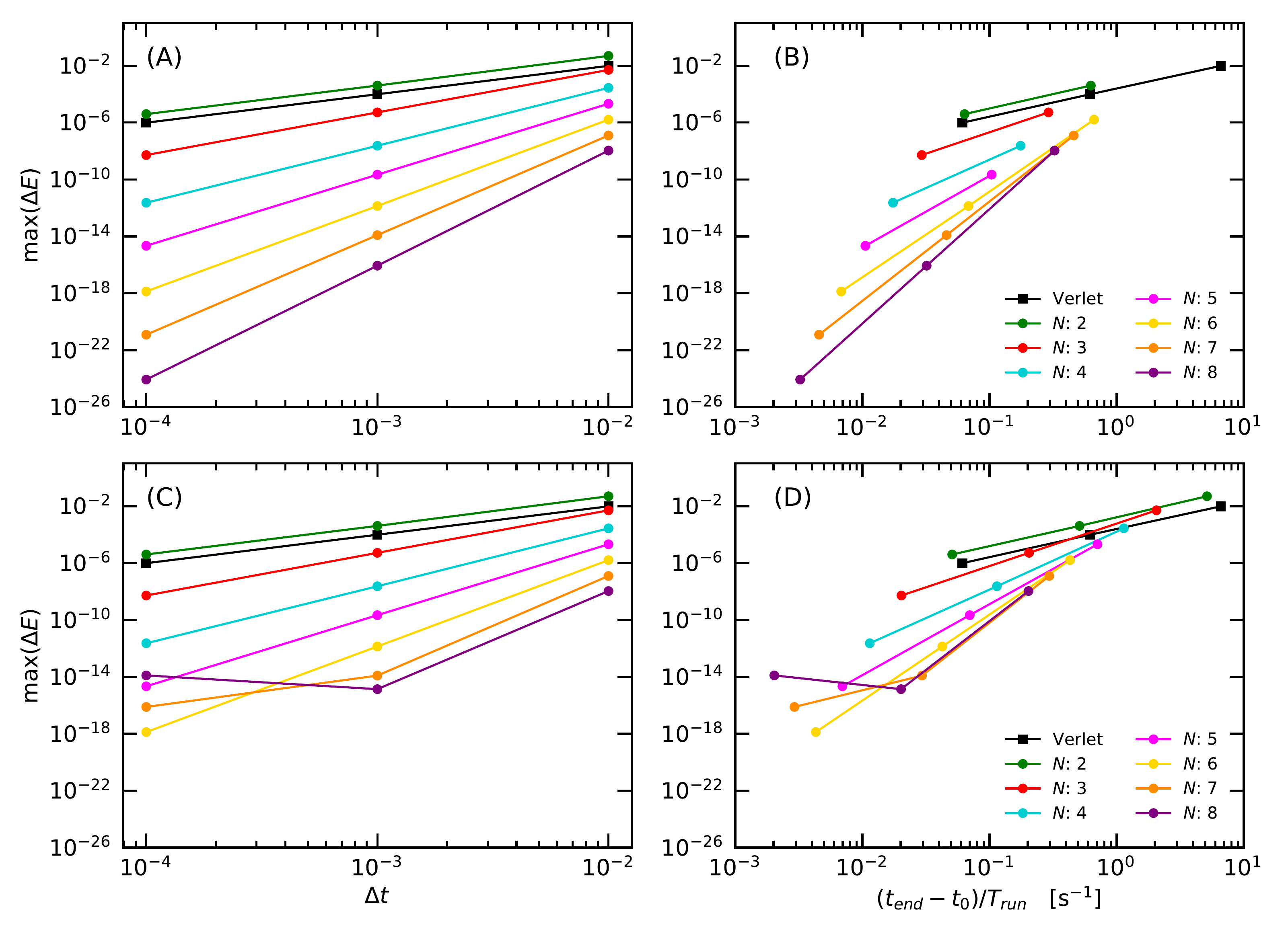}
\caption{Convergence with time step (A, C) and relation between accuracy and computational efficiency (B, D) for the Chebyshev (A, B) and Newton (C, D) polynomial propagators applied to a particle in Lennard-Jones potential. The convergence and the computational efficiency of the velocity Verlet integrator applied to the same problem are shown for comparison as black squares. The expressions were evaluated in precision of 30 digits using \texttt{mpmath}.}
\label{fig:lj-edrift-vs-tstep-prec=30}
\end{figure}

\begin{figure}
\includegraphics[width=\textwidth, trim=10 10 10 10, clip]{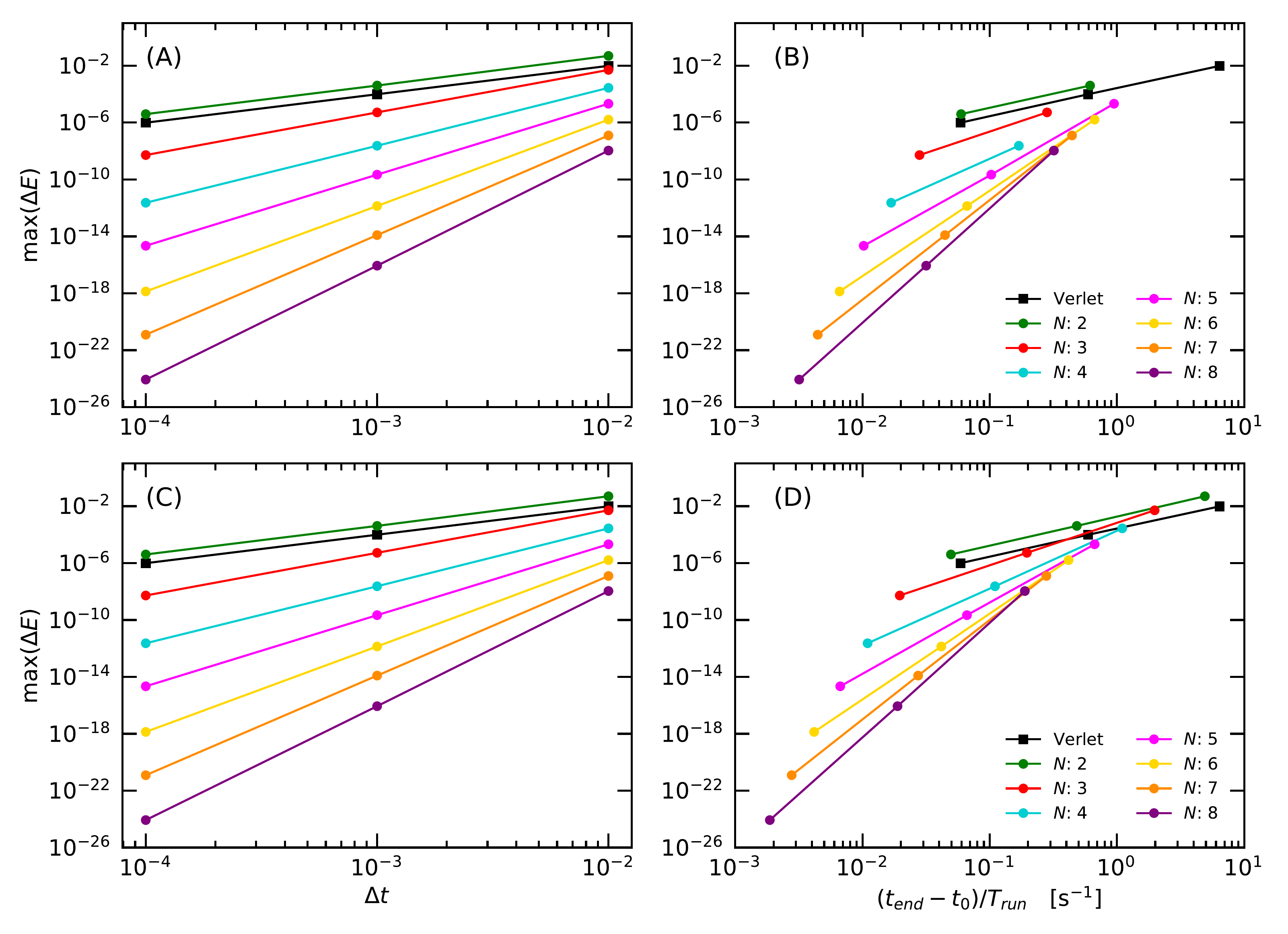}
\caption{Convergence with time step (A, C) and relation between accuracy and computational efficiency (B, D) for the Chebyshev (A, B) and Newton (C, D) polynomial propagators applied to a particle in Lennard-Jones potential. The  convergence and the computational efficiency of the velocity Verlet integrator applied to the same problem are shown for comparison as black squares. The expressions were evaluated in precision of 60 digits using \texttt{mpmath}.}
\label{fig:lj-edrift-vs-tstep-prec=60}
\end{figure}

\begin{figure}
\includegraphics[width=\textwidth, trim=10 10 10 10, clip]{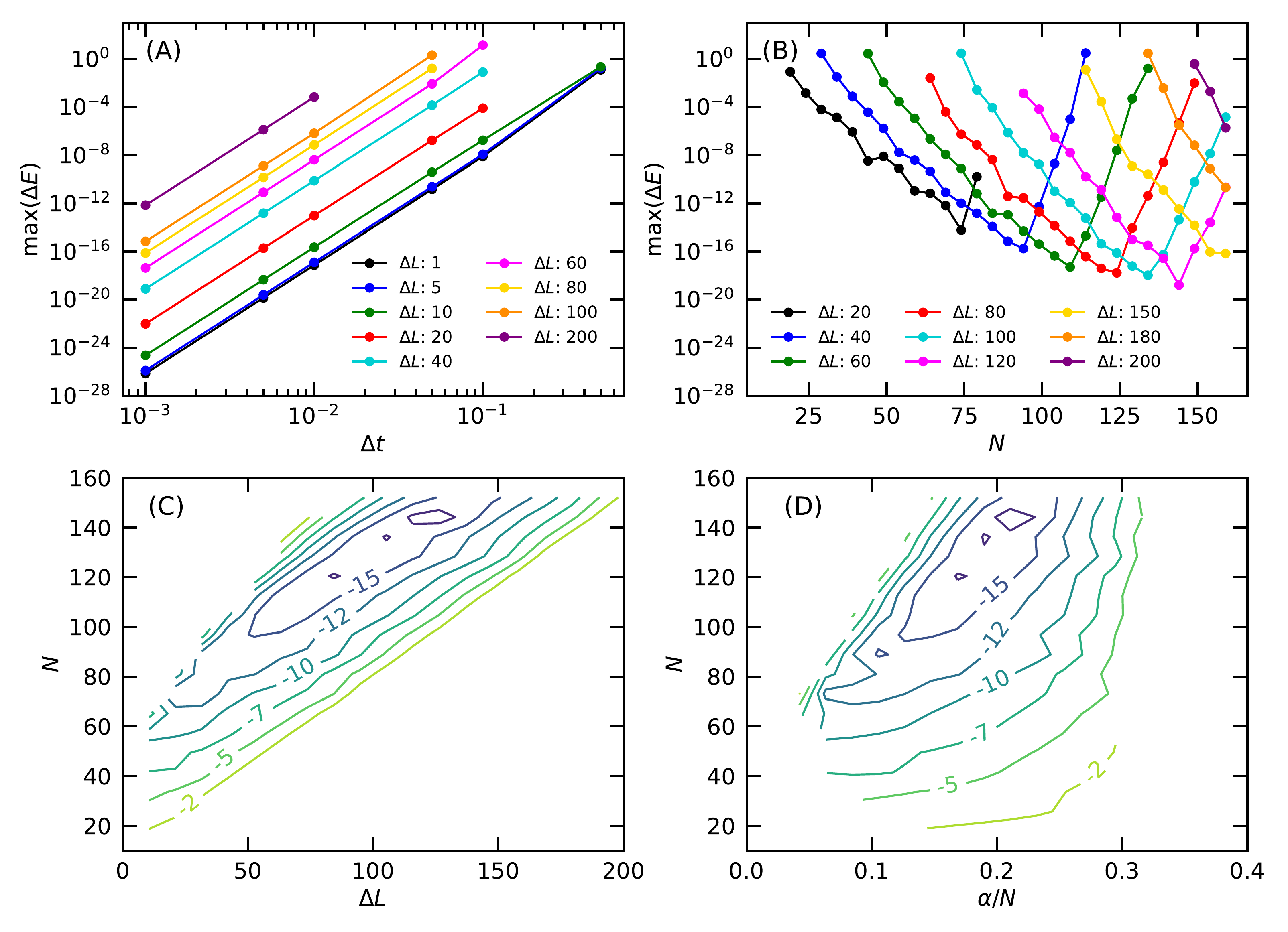}
\caption{The effect of the spectral width $\Delta L$ on the accuracy in terms of the maximum energy drift for the Newton polynomial propagator applied to the Morse oscillator. (A) The number of expansion terms is fixed to $N=10$. (B, C, D) The time step is fixed to $\Delta t = 0.5$. The expressions were evaluated in precision of 60 digits using \texttt{mpmath}.}
\label{fig:delta-L-newton-morse-prec=60}
\end{figure}

\begin{figure}
\includegraphics[width=\textwidth, trim=10 10 10 10, clip]{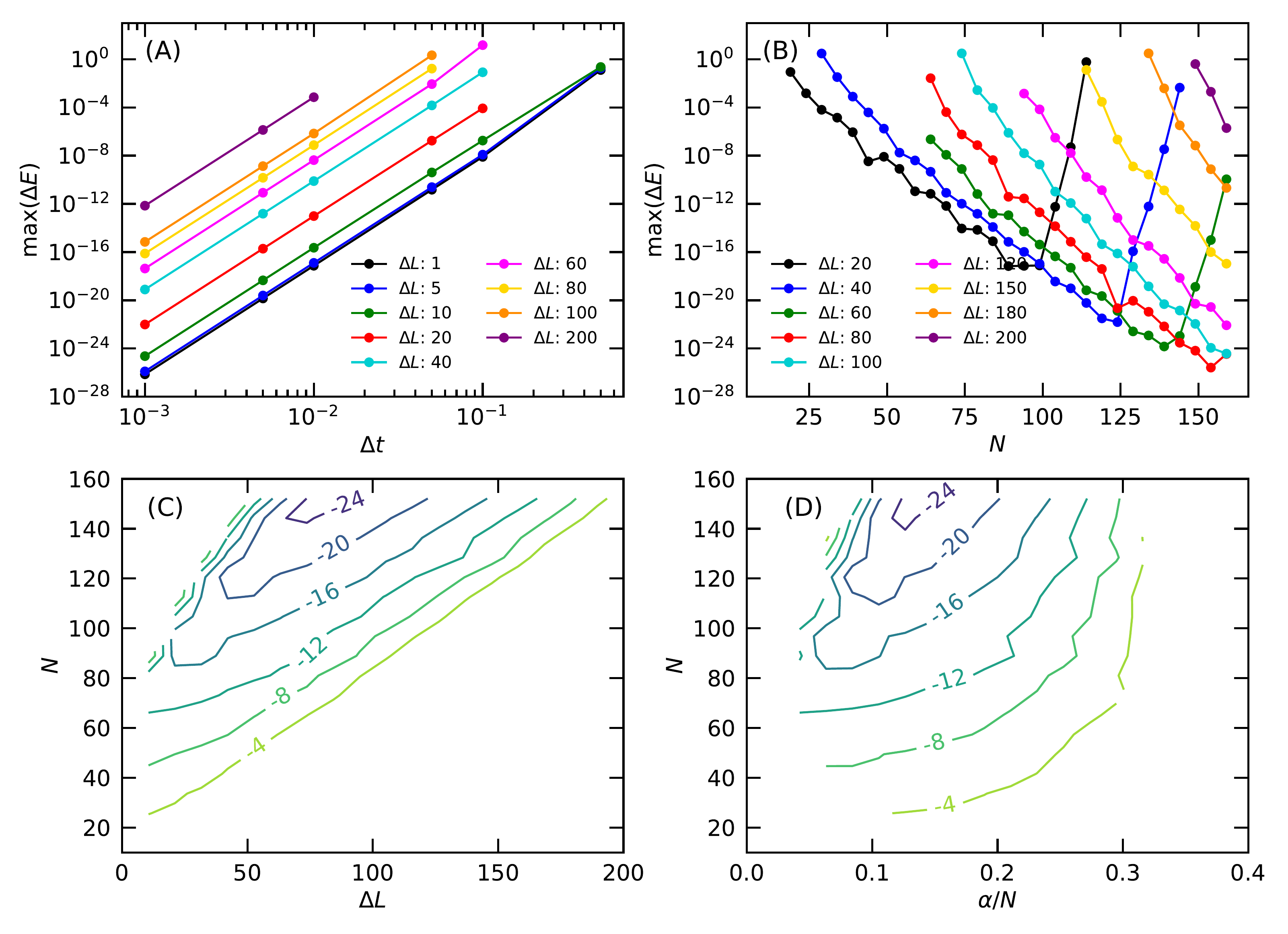}
\caption{The effect of the spectral width $\Delta L$ on the accuracy in terms of the maximum energy drift for the Newton polynomial propagator applied to the Morse oscillator. (A) The number of expansion terms is fixed to $N=10$. (B, C and D) The time step is fixed to $\Delta t = 0.5$. The expressions were evaluated in precision of 90 digits using \texttt{mpmath}.}
\label{fig:delta-L-newton-morse-prec=90}
\end{figure}

\begin{figure}
\includegraphics[width=\textwidth, trim=10 10 10 10, clip]{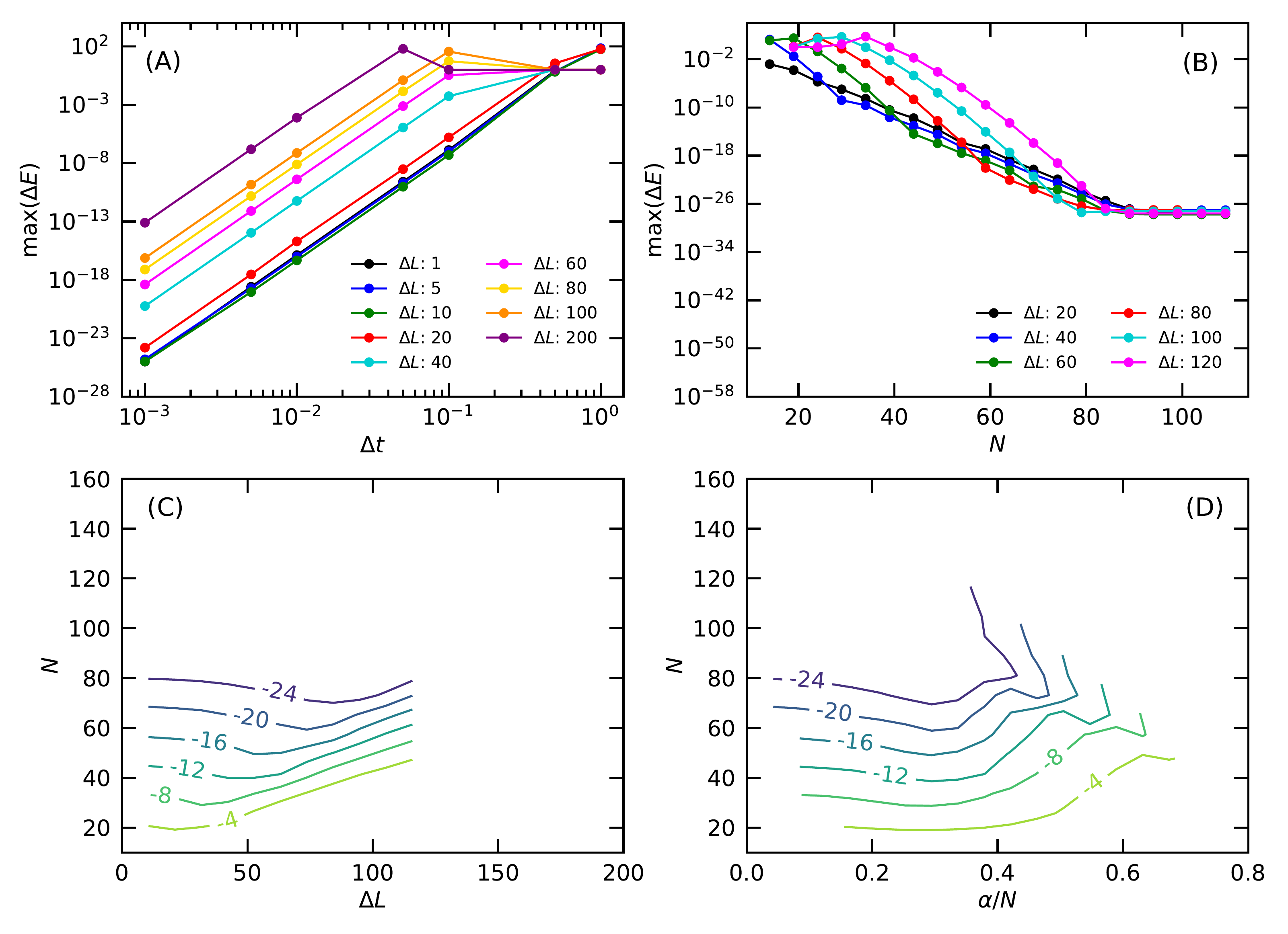}
\caption{The effect of the spectral width $\Delta L$ on the accuracy in terms of the maximum energy drift for the Chebyshev polynomial propagator applied to the anharmonic oscillator. (A) The number of expansion terms is fixed to $N=10$. (B, C and D) The time step is fixed to $\Delta t = 0.5$. The expressions were evaluated in precision of 30 digits using \texttt{mpmath}.}
\label{fig:delta-L-chebyshev-anharmonic-prec=30}
\end{figure}

\begin{figure}
\includegraphics[width=\textwidth, trim=10 10 10 10, clip]{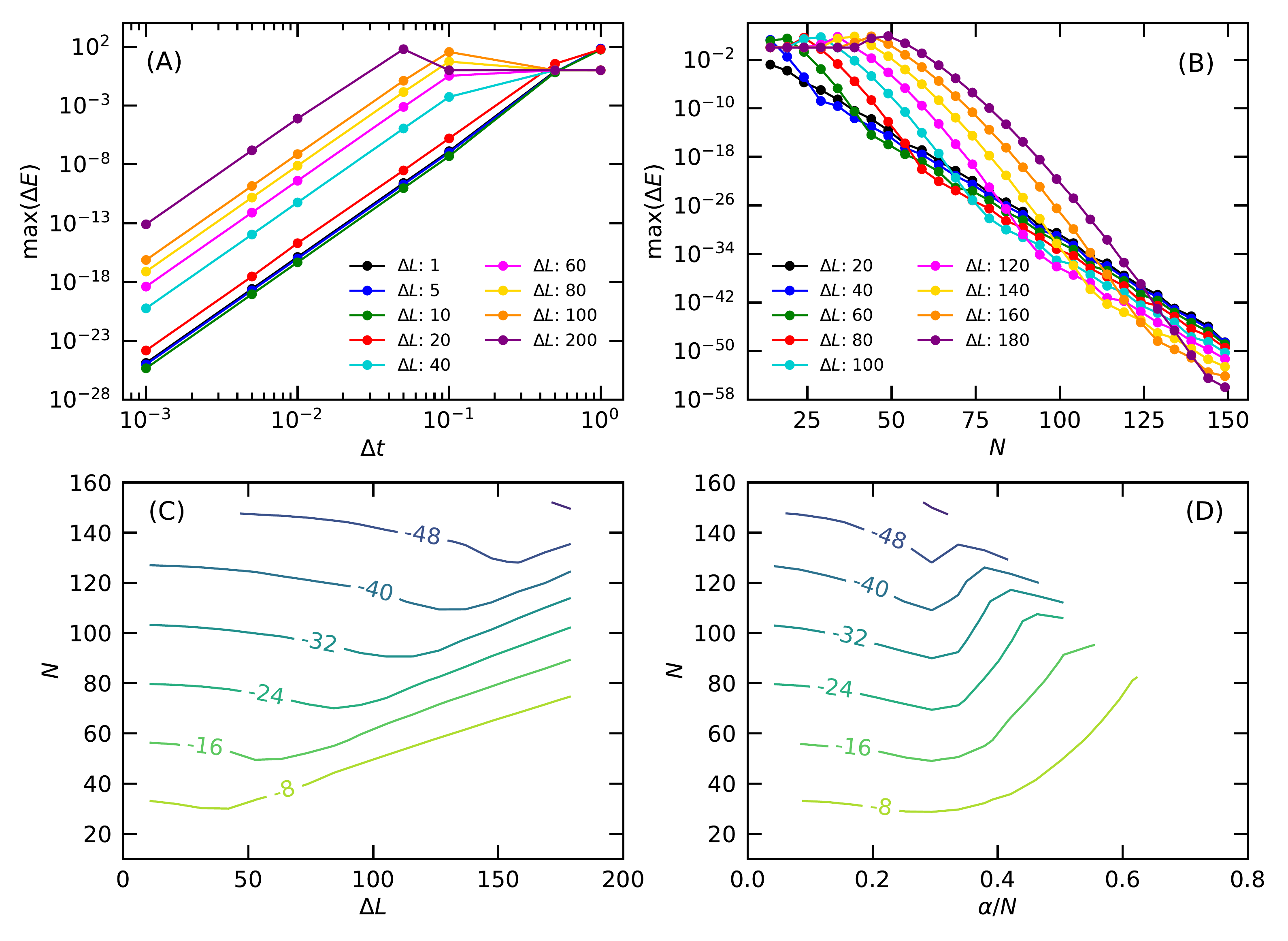}
\caption{The effect of the spectral width $\Delta L$ on the accuracy in terms of the maximum energy drift for the Chebyshev polynomial propagator applied to the anharmonic oscillator. (A) The number of expansion terms is fixed to $N=10$. (B, C and D) The time step is fixed to $\Delta t = 0.5$. The expressions were evaluated in precision of 60 digits using \texttt{mpmath}.}
\label{fig:delta-L-chebyshev-anharmonic-prec=60}
\end{figure}

\begin{figure}
\includegraphics[width=\textwidth, trim=10 10 10 10, clip]{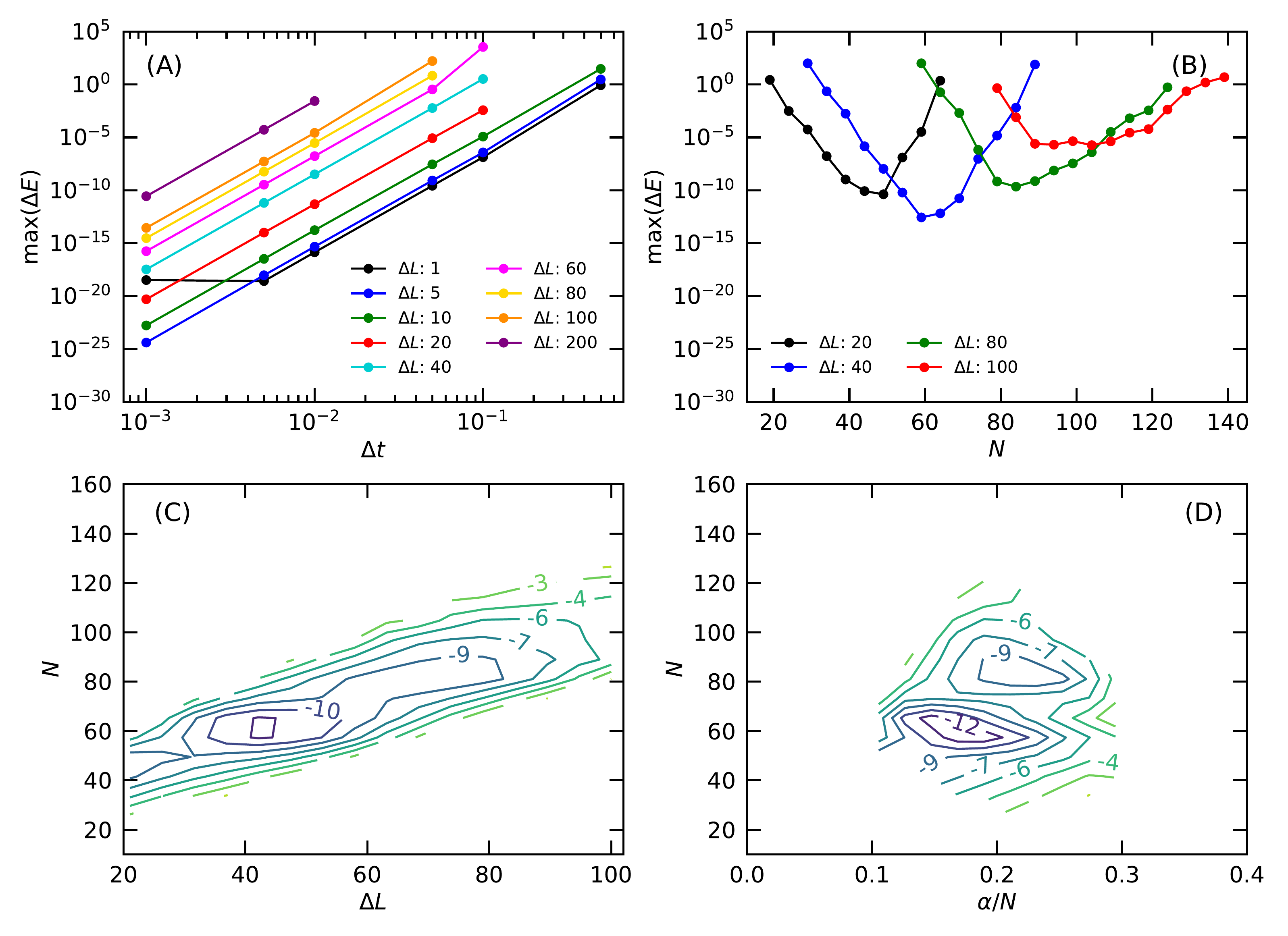}
\caption{The effect of the spectral width $\Delta L$ on the accuracy in terms of the maximum energy drift for the Newton polynomial propagator applied to the anharmonic oscillator. (A) The number of expansion terms is fixed to $N=10$. (B, C and D) The time step is fixed to $\Delta t = 0.5$. The expressions were evaluated in precision of 30 digits using \texttt{mpmath}.}
\label{fig:delta-L-newton-anharmonic-prec=30}
\end{figure}

\begin{figure}
\includegraphics[width=\textwidth, trim=10 10 10 10, clip]{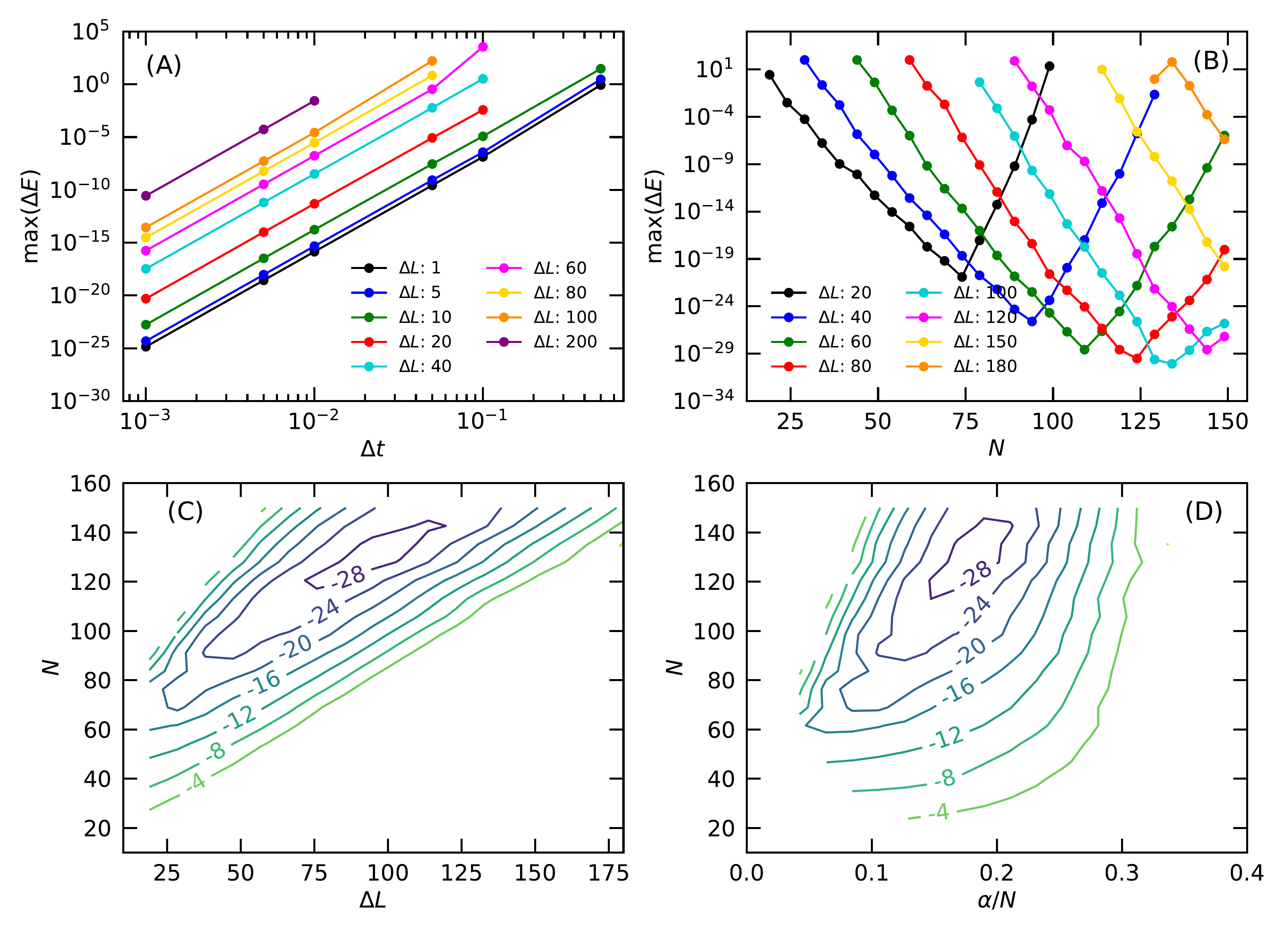}
\caption{The effect of the spectral width $\Delta L$ on the accuracy in terms of the maximum energy drift for the Newton polynomial propagator applied to the anharmonic oscillator. (A) The number of expansion terms is fixed to $N=10$. (B, C, and D) The time step fixed to $\Delta t = 0.5$. The expressions were evaluated in precision of 60 digits using \texttt{mpmath}.}
\label{fig:delta-L-newton-anharmonic-prec=60}
\end{figure}

\begin{figure}
\includegraphics[width=\textwidth, trim=10 10 10 10, clip]{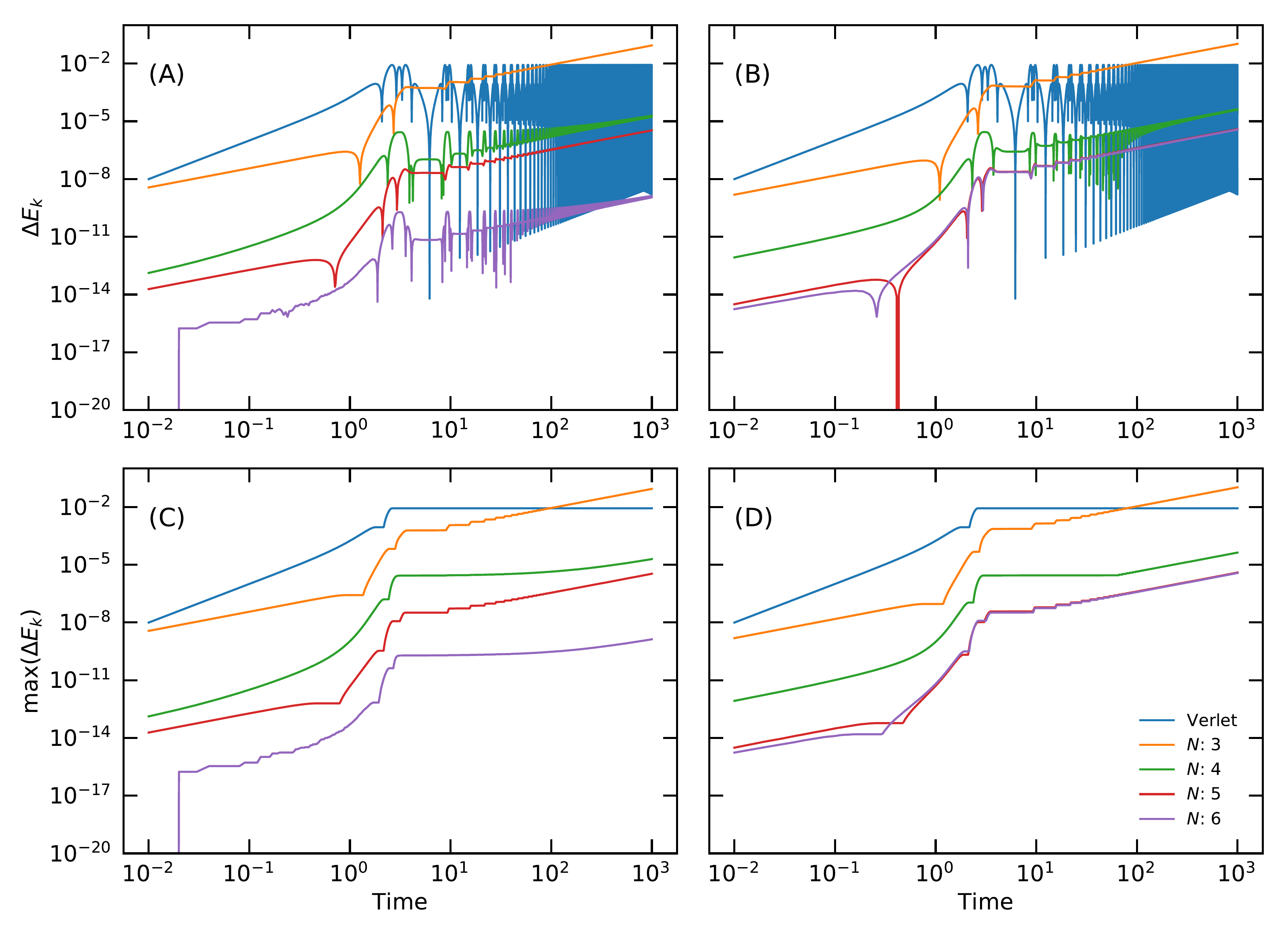}
\caption{Accuracy of the Chebyshev (A, C) and Newton (B, D) polynomial propagators measured by the energy drift (A, B) for the anharmonic oscillator and the running maximum of the energy drift (C, D) for a long period of integration with $10^5$ time steps with size $\Delta t=0.01$. For comparison, the energy drift and the running maximum of the energy drift were measured with the velocity Verlet method with the same time step size. The expressions were numerically evaluated in double precision. }
\label{fig:long-time-anharmonic-dp}
\end{figure}

\begin{figure}
\includegraphics[width=\textwidth, trim=10 10 10 10, clip]{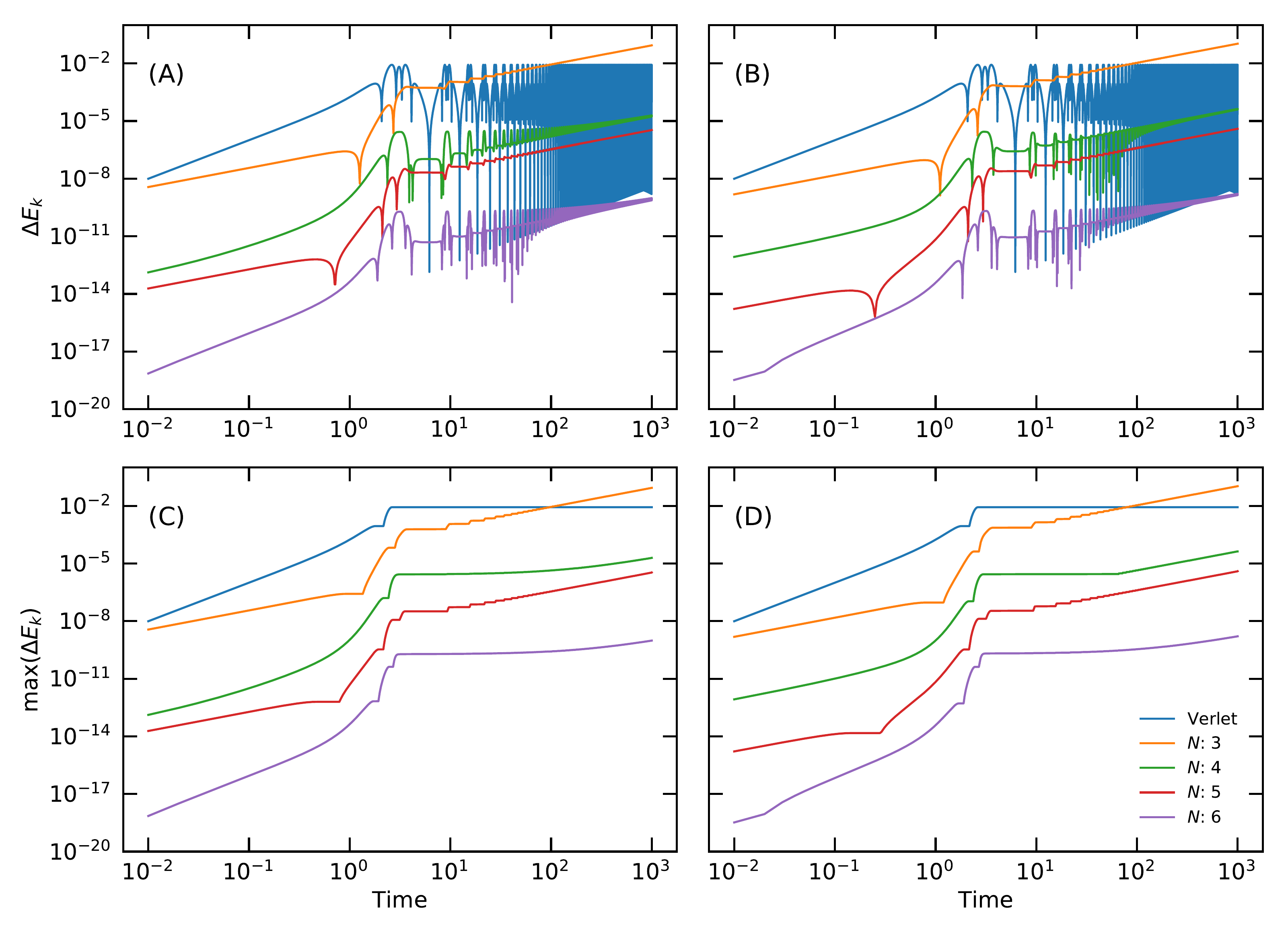}
\caption{Accuracy of the Chebyshev (A, C) and Newton (B, D) polynomial propagators measured by the energy drift (A, B) for the anharmonic oscillator and the running maximum of the energy drift (C, D) for a long period of integration with $10^5$ time steps with size $\Delta t=0.01$. For comparison, the energy drift and the running maximum of the energy drift were measured with the velocity Verlet method with the same time step size. The expressions were evaluated in precision of 30 digits using \texttt{mpmath}.}
\label{fig:long-time-anharmonic-prec=30}
\end{figure}

\begin{figure}
\includegraphics[width=\textwidth, trim=10 10 10 10, clip]{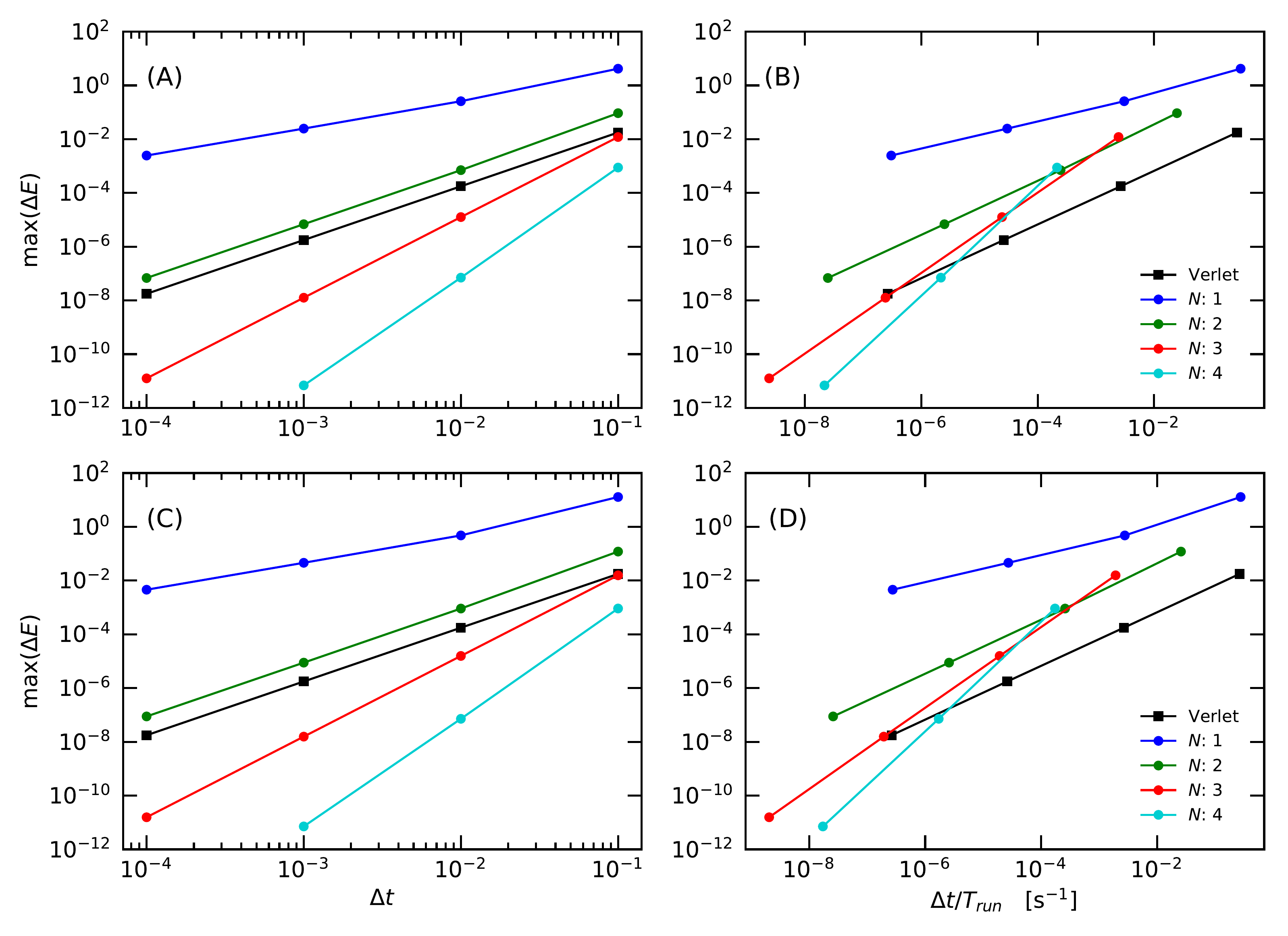}
\caption{Convergence with time step (A, C) and relation between accuracy and computational efficiency (B, D) for the Chebyshev (A, B) and Newton (C, D) polynomial propagators applied to two particles interacting through a Morse potential. The  convergence and the computational efficiency of the velocity Verlet integrator applied to the same problem are shown for comparison as black squares. The expressions were compiled with the \texttt{lambdify} tool and evaluated in precision of 30 digits with the \texttt{mpmath} backend.}
\label{fig:2p-morse-edrift-vs-tstep-prec=30}
\end{figure}

\begin{figure}
\includegraphics[width=\textwidth, trim=10 10 10 10, clip]{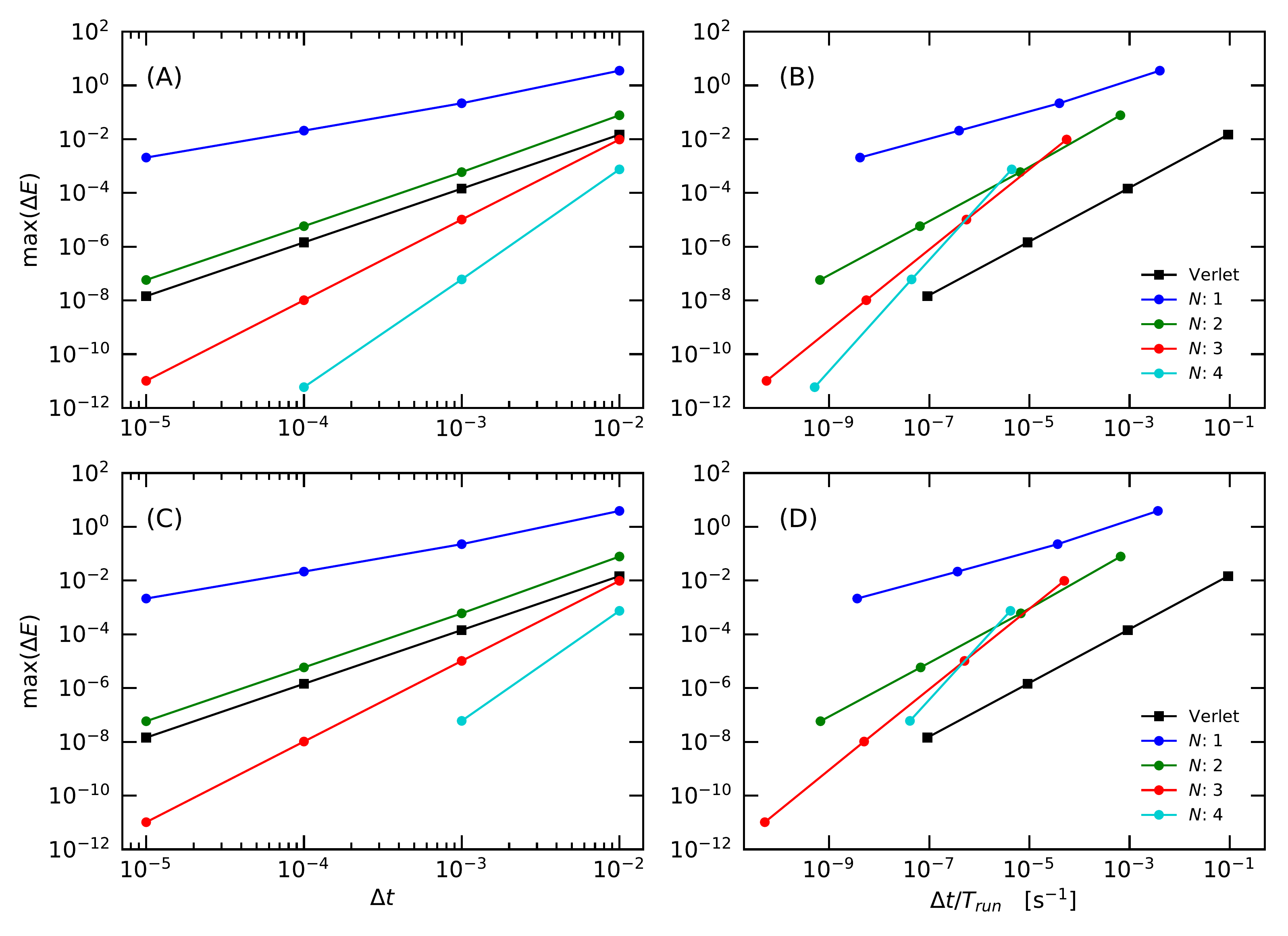}
\caption{Convergence with time step (A, C) and relation between accuracy and computational efficiency (B, D) for the Chebyshev (A, B) and Newton (C, D) polynomial propagators applied to two particles interacting through a Lennard-Jones potential. The  convergence and the computational efficiency of the velocity Verlet integrator applied to the same problem are shown for comparison as black squares. The expressions were compiled with the \texttt{lambdify} tool and evaluated in precision of 30 digits with the \texttt{mpmath} backend.}
\label{fig:2p-lj-edrift-vs-tstep-prec=30}
\end{figure}

\begin{figure}
\includegraphics[width=\textwidth, trim=10 10 10 10, clip]{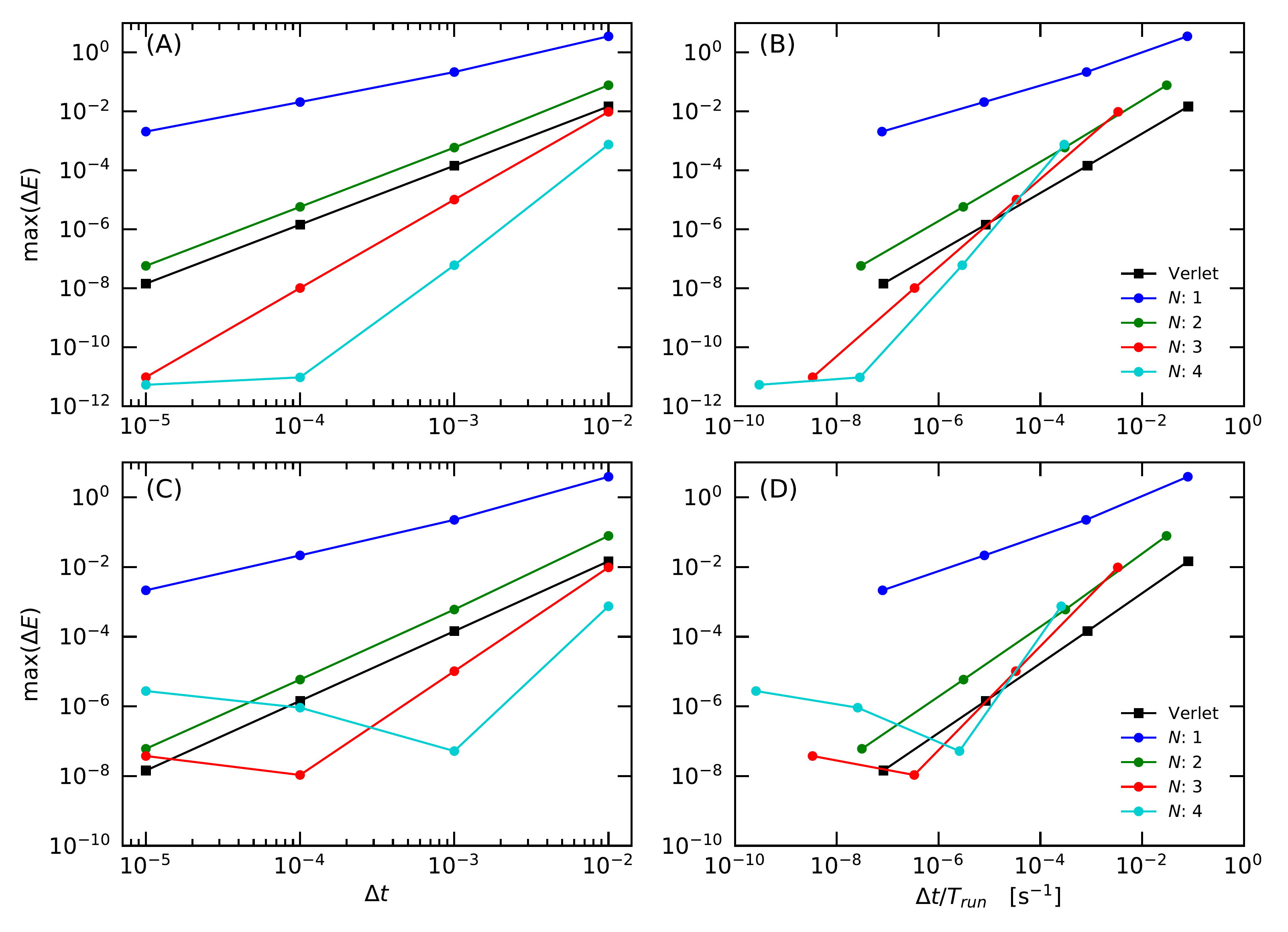}
\caption{Convergence with time step (A, C) and relation between accuracy and computational efficiency (B, D) for the Chebyshev (A, B) and Newton (C, D) polynomial propagators applied to two particles interacting through a Lennard-Jones potential. The  convergence and the computational efficiency of the velocity Verlet integrator applied to the same problem are shown for comparison as black squares. The expressions were compiled with \texttt{lambdify} and evaluated in double precision with the \texttt{numpy} backend.}
\label{fig:2p-lj-edrift-vs-tstep-dp-lambdify}
\end{figure}

\begin{figure}
\includegraphics[width=\textwidth, trim=10 10 10 10, clip]{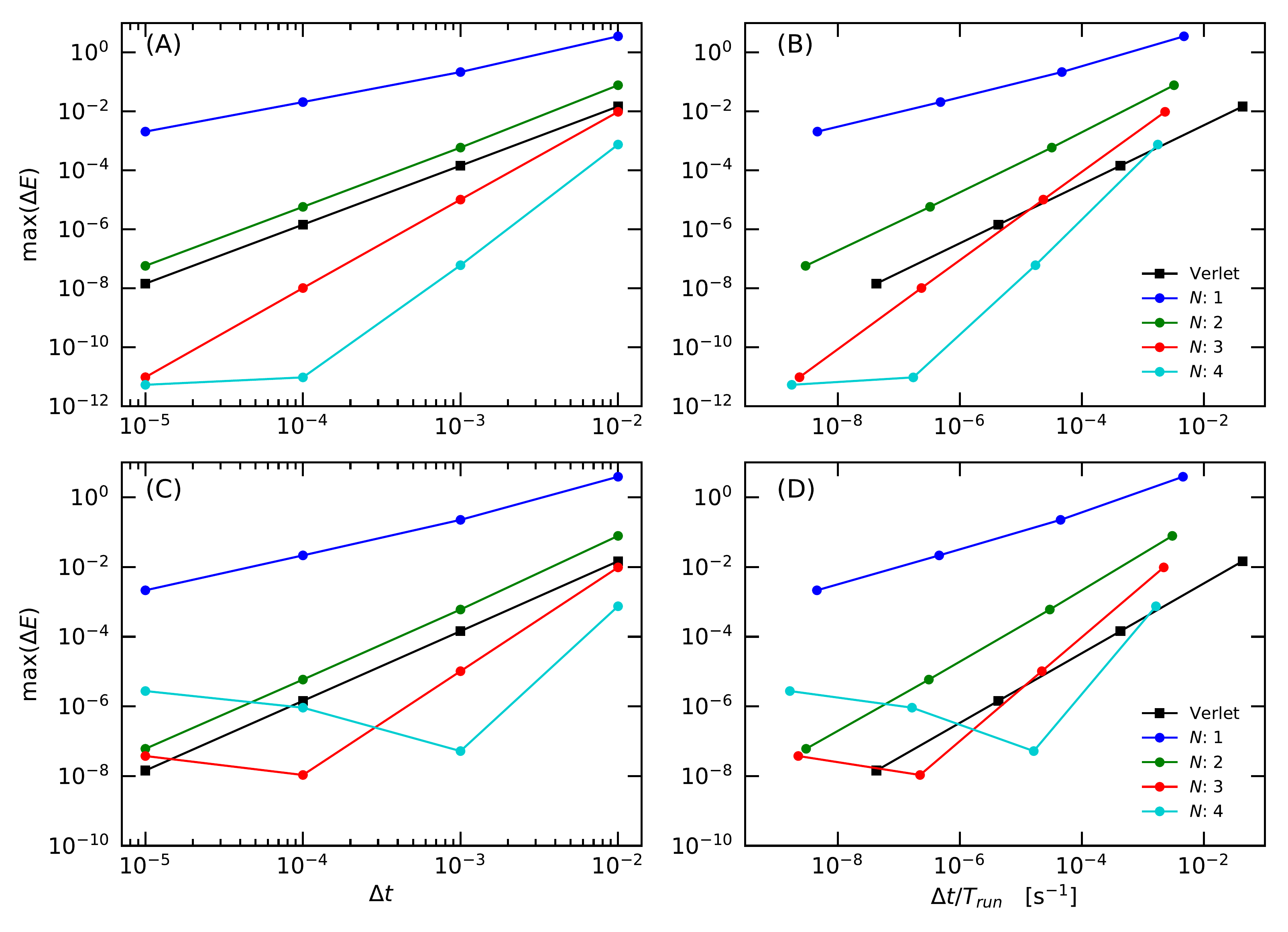}
\caption{Convergence with time step (A, C) and relation between accuracy and computational efficiency (B, D) for the Chebyshev (A, B) and Newton (C, D) polynomial propagators applied to two particles interacting through a Lennard-Jones potential. The convergence and the computational efficiency of the velocity Verlet integrator applied to the same problem are shown for comparison as black squares. The expressions were compiled and numerically evaluated with \texttt{theano} in double precision.}
\label{fig:2p-lj-edrift-vs-tstep-dp-theano}
\end{figure}

\begin{figure}
\includegraphics[width=\textwidth, trim=10 10 10 10, clip]{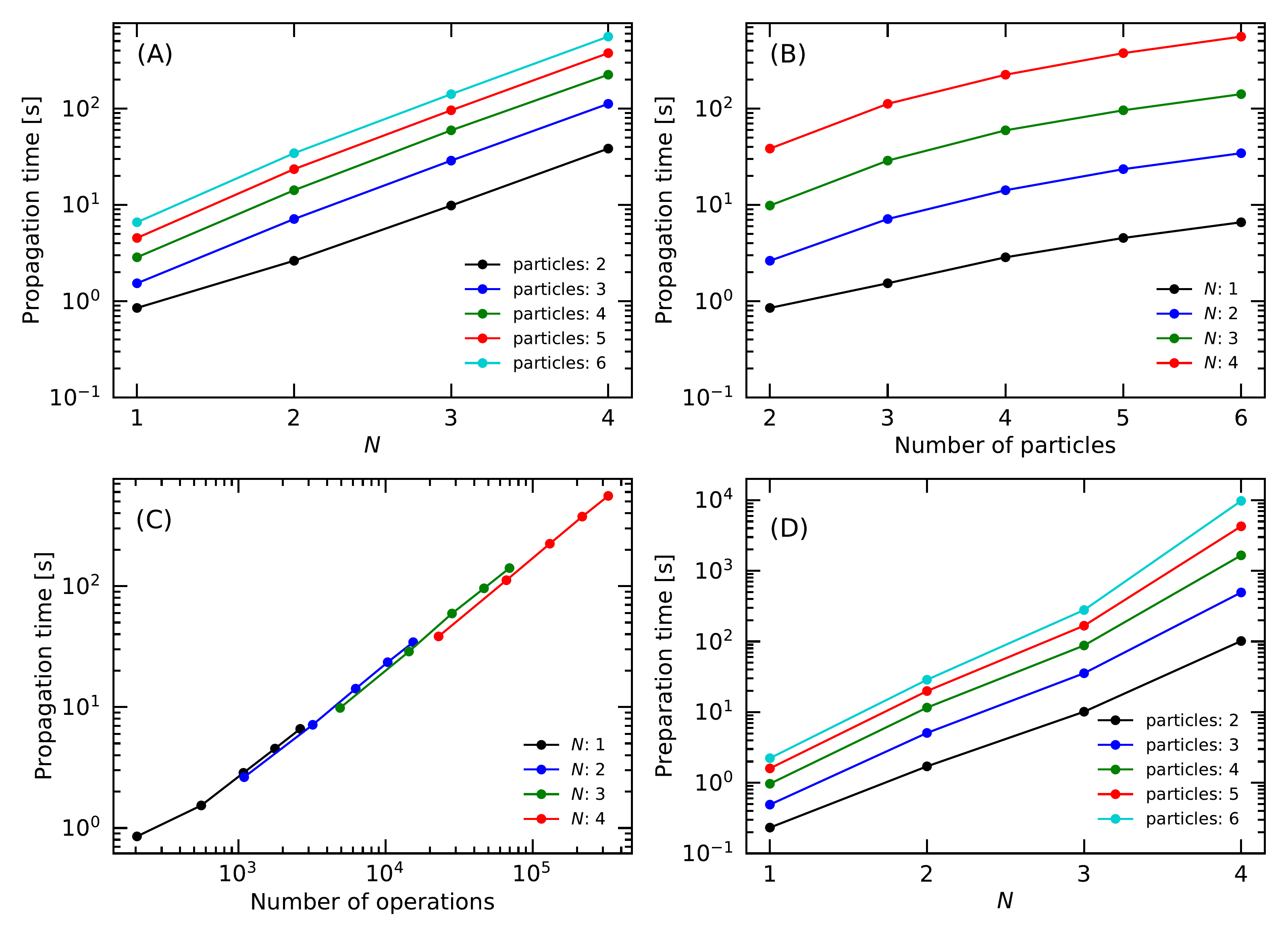}
\caption{Performance of the Newton propagator in a simulation of Ar atoms interacting through a Lennard-Jones potential measured for with time step size of $\Delta t = 0.001$. The \texttt{lambdify} tool was used to compile the expressions that were evaluated with the \texttt{numpy} backend in double precision.}
\label{fig:many-particles-newton}
\end{figure}